\documentclass[prd,twocolumn,nofootinbib]{revtex4-2}

\usepackage{amsfonts}
\usepackage{yfonts}

\expandafter\let\csname equation*\endcsname\relax
\expandafter\let\csname endequation*\endcsname\relax
\usepackage{amsmath}

\usepackage{amssymb}
\usepackage{bm}
\usepackage{dcolumn}
\usepackage{graphicx}
\usepackage{graphics}
\graphicspath{{figures/}}
\usepackage[utf8]{inputenc}
\usepackage{latexsym}
\usepackage{rotating}
\usepackage{hyperref}
\usepackage[all]{hypcap} 
\usepackage{xspace} 
\usepackage[usenames,dvipsnames]{color}
\usepackage{mathrsfs}
\usepackage{subfigure}
\usepackage{xcolor}
\usepackage{soul}

\usepackage{ulem}
\usepackage{outlines}
\usepackage{enumitem}
\setenumerate[1]{label=\Roman*.}
\setenumerate[2]{label=\Alph*.}
\setenumerate[3]{label=\roman*.}
\setenumerate[4]{label=\alph*.}

\definecolor {darkgreen}{rgb}{0.2,0.7,0.2}

\usepackage{etoolbox}

\makeatletter
\def\@mkboth#1#2{}
\newlength\appendixwidth
\preto\appendix{\addtocontents{toc}{\protect\patchl@section}}
\newcommand{\patchl@section}{%
  \settowidth{\appendixwidth}{\textbf{Appendix }}%
  \addtolength{\appendixwidth}{1.5em}%
  \patchcmd{\l@section}{1.5em}{\appendixwidth}{}{\ddt}%
}
\makeatother

\DeclareMathOperator{\diag}{diag}
\DeclareMathOperator{\Res}{Res}
\newcommand{\dee}{\operatorname{d\!}{}}

\newcommand{\bxi}{\bm{\xi}}
\newcommand{\Utidal}{U_{\textrm{tidal}}}

\newcommand{\STFu}[2]{#1^{\left< #2 \right>}}

\newcommand{\Yc}[1]{\mathcal{Y}^{*\left< #1 \right>}}

\newcommand{\tp}{t_\textrm{p}}
\newcommand{\MB}{M_{\rm B}}

\newcommand{\nn}{\nonumber}

\begin{document}
\title{Neutron Stars in the Effective Fly-By Framework:\\$f$-Mode Re-summation}

\author{J. Nijaid Arredondo}
\affiliation{Department of Astrophysical Sciences, Princeton University, Princeton, NJ, 08544, USA}
\author{Nicholas Loutrel}
\affiliation{Department of Physics, Princeton University, Princeton, NJ, 08544, USA}

\date{\today}

\begin{abstract} 
  Eccentric compact binaries pose not only a challenge for
  gravitational wave detectors,
  but also provide a probe into the nuclear equation of state if one of the objects is a neutron star.
  At the short pericenter passage, tidal interactions excite f-modes
  on the star, which in turn emit their own gravitational waves.
  We derive an analytic waveform for these stellar oscillations
  within the effective fly-by framework,
  modeling the emission to leading post-Newtonian order. At this order, the f-mode response can be written in a Fourier decomposition in terms of orbital harmonics, with the amplitudes of each harmonic depending on Hansen coefficients.
  Re-summing the harmonics of the f-mode results
  in a simple decaying harmonic oscillator, with the amplitude now determined by a Hansen coefficient of complex harmonic number. We compute the match between the re-summed f-mode and numerical integrations of the tidal response, and find ${\cal{M}} > 0.98$ for systems with high orbital eccentricity $(e > 0.8)$ and low semi-latus rectum $(p < 15 M)$ for three equations of state.
  We further compare our model to modes generated from subsequent pericenter passages under the effect of radiation reaction, and develop an accurate model to time pericenter passages. We show how the timing model can be used to specify initial conditions to accurately track the f-mode excitation across multiple pericenter passages.
\end{abstract}

\maketitle


\newpage
\section{Introduction}

It is now five years since the
Laser Interferometer Gravitational-wave Observatory (LIGO)~\cite{TheLIGOScientific:2014jea}
first detected the gravitational waves of merging binary black holes (BBHs) \cite{Abbott2016}.
Since then, the LIGO and Virgo~\cite{Caron:1997hu} Scientific Collaborations have in total
detected the coalescence of fifty pairs of
compact objects,
including black holes (BHs) and neutron stars (NSs) \cite{Abbott2020y}.
LIGO and Virgo initiated an era of multi-messenger astronomy with the detection
of its first NS--NS binary \cite{Abbott2017a},
as it was accompanied by the electromagnetic observation of its kilonova \cite{Abbott2017d}.
This has allowed for quantitative tests of the structure of these NSs
probed by the strong gravity during merger \cite{Abbott2018},
constraining the equation of state (EOS) of degenerate matter
with measurement of the stars' masses, radii, and tidal deformabilities \cite{Raithel2019}.
With this data and upcoming observations,
gravitational wave (GW) astronomy has opened up promising avenues
towards probing strong gravity and the equation of state of nuclear matter.

GW astronomy is yet in its infancy,
and its ability to characterize detections has already been
pushed into uncharted territory.
Only 4 cycles were captured of LIGO-Virgo detection GW190521,
leading to uncertainties in its interpretation \cite{Abbott2020mjq}.
Besides its primary mass lying in the pair-instability mass gap,
making it an outlier from previous LIGO and Virgo events,
its short duration makes it difficult to determine the precise interpretation of the signal, specifically quasi-circular spin precessing versus head-on collision.
Follow up studies performing model selection have found that models with non-zero eccentricity $(e>0.1)$ are favored over quasi-circular precessing models~\cite{Romero2020}, while high eccentricity $(e\sim0.7)$ precessing waveforms are even more favored~\cite{Gayathri2020}. However, the uncertainties between quasi-circular, precessing binary models
and eccentric models hamper this
interpretation from being definitive \cite{Romero2020, CalderonBustillo:2020odh}.

A conclusive measurement
of a merging binary with non-zero eccentricity
would be a strong indicator of dynamical formation,
tracing the formation channel of the merging components
in dense stellar environments.
LIGO-Virgo's detections have so far been consistent with the
rates expected from isolated binary evolution,
although this may be due to the large uncertainties
in spin measurements and population models \cite{Abbott2020gyp}.
This formation channel is expected to produce orbits in which
the components' spins are predominantly aligned
with the orbital angular momentum
and are circularized by the time they enter LIGO's frequency band.
Such assumptions will unfortunately exclude dynamical channels
in which binaries merge with significant eccentricity or
misaligned spins \cite{Rodriguez2016}.
These channels are sourced by a diversity of dynamical processes,
such as GW captures between two unbound objects \cite{Rodriguez2018pss},
binary-single interactions including
Kozai--Lidov cycles \cite{Antonini2012,Antonini2016}
and exchanges \cite{Samsing2017xmd},
and hierarchical mergers \cite{Samsing2013kua},
which is a possible formation channel for GW190521 \cite{Liu2020gif}.
These processes have the potential to create
eccentric binaries in dense stellar environments
whose GWs may be detectable with a variety of detectors, including decihertz detectors, LISA, and LIGO-Virgo
\cite{Kremer2018cir,Samsing2019dtb,Sedda2019uro}.
It is estimated that 30\% of the inner binaries in triple systems formed in globular clusters may have $e>0.1$ when they enter the LIGO band thanks to Kozai--Lidov cycles \cite{Brown:2009ng},
and 90\% of stellar-mass BH binaries formed by scattering in galactic cores may even have $e>0.9$ \cite{Barta:2018tyg}.

Dynamical formation channels are not only expected to produce BBHs, but also BH--NS binaries and binary neutron stars (BNSs).
The BH--NS merger rates in dense environments have only been loosely constrained
and vary depending on their formation channel.
For the more promising channels,
the rate for binary-single interactions has been estimated to be as
high as 0.25 Gpc$^{-3}$ yr$^{-1}$ in globular clusters \cite{Sedda2020wzl};
for Kozai--Lidov to be as high as 0.33 Gpc$^{-3}$ yr$^{-1}$
in galactic nuclei \cite{Stephan2019fhf,Hoang2020};
and for the combination of binary evolution and dynamical exchanges in
young star clusters to be as high as $\sim 28$ Gpc$^{-3}$ yr$^{-1}$ \cite{Rastello2020sru},
although this estimate assumes low natal NS kicks
and high densities that may not be representative of all young star clusters
(a more conservative upper limit has been found to be
$\sim 10^{-3}$ Gpc$^{-3}$ yr$^{-1}$;
see \cite{Hoang2020}).
The latter category is especially interesting as almost all of the
dynamically assembled BH--NSs are ejected from their cluster,
thus merging in the field and possibly mixing with binaries that did not form in clusters.
These are all a fraction of LIGO-Virgo's empirical upper BH--NS merger rate
of 610 Gpc$^{-3}$ yr$^{-1}$ \cite{Abbott2018mvr},
which is estimated to be dominated by isolated binaries in the field.
However, triple systems in the field may form eccentric binaries
within LIGO;
the rate of BH--NS mergers formed from these triples
is estimated to be $10^{-3} - 19$ Gpc$^{-3}$ yr$^{-1}$ \cite{Fragione2019zhm}.
Understanding these channels and being able to identify their signals
will be crucial as detections increase.

The method of power stacking has been proposed as a means of detecting the GWs from eccentric binaries~\cite{Tai2014}.
While quasi-circular orbits emit sinusoidal GWs,
eccentric orbits emit bursts of radiation
that are increasingly localized around pericenter
the more eccentric the orbit.
By searching for these bursts as excess power in the detector data
and stacking them,
detectors can be more sensitive to signals from eccentric binaries.
To distinguish these from other sources of uncorrelated excess power
such as glitches \cite{Cabero2019,Abbott2016o},
a timing model that can physically correlate a sequence of bursts
would be needed.
Such a model has so far only been developed in the post-Newtonian (PN)
formalism for point masses \cite{Loutrel2017b}.

However, to detect and reconstruct a signal,
the method of choice is match filtering with template waveforms \cite{Abbott2020y}.
Complications that arise from non-circular effects,
such as orbital precession and the timescales involved
in computing multiple eccentric orbits
make quasi-circular waveforms simpler to produce and analyze.
Some of these complications arise in numerical simulations
that model the evolution of the spacetime with full general relativity,
as accurately capturing the GWs at pericenter passage
requires resolving timescales shorter
than the period of an orbit.
For eccentric orbits,
these timescales are much more disparate,
with the time spent at pericenter being orders of magnitude smaller.
Yet numerical simulations are key to accurately modeling the waveforms
used in match filtering \cite{Dietrich2017aum,Dietrich2018uni}.

While LIGO has ruled out detections with significant eccentricity in its
first two observing runs \cite{Abbott2019owp},
it is recognized that quasi-circular waveforms
lose sensitivity when matched against eccentric signals,
in particular if the orbit is of low mass \cite{Martel1999}.
Quantitatively,
these models lose significant sensitivity for $e \sim 0.07$ \cite{Nitz2020}
and may miss BNS with $e \geq 0.02$ \cite{Moore2019vjj}.
Significant efforts have gone into modeling the GW emission from eccentric binaries in recent years,
with PN models currently available up to 3PN order and up to $e\sim 0.6$ \cite{Moore:2019xkm}, as well as effective one-body
models at moderate~\cite{Cao:2017ndf, Chiaramello:2020ehz} and high~\cite{Nagar:2020xsk} eccentricities. So far, these models have focused on BBHs and do not
include the tidal effects important for studying neutron stars.
This is an aspect that needs addressing to keep systematic errors of models below the statistical errors of NS--NS inspirals.
The systematic errors can supersede the statistical starting at the 3.5PN order for LIGO if eccentricity is not considered with tidal effects, biasing the recovered source parameters from a detection \cite{Favata:2013rwa}.

Modeling eccentric systems probes properties beyond the formation channel
of an orbit -- it also elucidates the structure of its components.
The close passage between a BH and a NS at pericenter
causes strong tides that exchanges energy from the orbit
into oscillations on the star \cite{Press1977}.
BH--NS binaries will constitute a fraction of eccentric mergers,
and their detection may further constrain NS models.
Some numerical studies of the inspiral have been performed
\cite{Gold:2011df,East2015vix,East2011aa},
quantifying their GWs and ejecta that can be analyzed for
multi-messenger signals \cite{Papenfort:2018bjk}, including gamma-ray bursts \cite{Stephens2011as}.
In particular, the GW signal carries the signature of the
star's fundamental modes (f-modes) \cite{Lindblom:1983,East2011xa}.
These oscillations are EOS-dependent,
thus identifying the structure of the NS \cite{Thorne1967,Kokkotas1999,Pratten2019sed}.
Their frequencies and damping times have also
been calculated from various EOSs to derive
universal relations that describe them as functions of a star's
mass and radius \cite{Andersson1998,Lioutas2017xtn,Lioutas2020vzi}.

Taking steps towards waveform models of eccentric binaries with NSs,
analytic PN treatments have focused on the coupling of the orbital evolution
to f-modes \cite{Parisi2018,Yang2018}.
The latter found for a polytropic star that the mode energy
can be on the order of the emitted GW energy for very
close passages, in agreement with numerical simulations.
The phase shift in the waveform has been estimated to leading order
in a small-eccentricity approximation,
finding that to measure it requires the sensitivity
of third-generation detectors,
or at least a LIGO configuration tuned to high frequencies \cite{Yang2019}.
Numerical integrations of the PN equations have also shown
that the energy deposited into f-modes can grow chaotically for a
BNS binary and reach $\sim 5\%$ of the NS binding energy,
given an initial separation on the order of the NS radius
(see Fig. 3 of \cite{Vick2019}).
The phase shift of the GW can reach the tens of radians before merger,
demonstrating the effect of the f-modes on hastening the collision.

The effects of eccentricity and EOS on waveforms
have thus shown exciting results.
Detectors that are more sensitive to them across different frequency bands
are on the horizon,
such as the space-borne LISA \cite{Robson2018}
and the third-generation ground detectors Cosmic Explorer (CE) \cite{Reitze2019}
and Einstein Telescope (ET) \cite{Punturo2010}.
Upon realization, the third-generation is expected to reduce LIGO uncertainties
on the EOS by an order of magnitude \cite{Hinderer2010}.
Eccentric waveforms will need to be available by then
to accurately characterize formation channels and the structure of NSs.

\subsection{Executive Summary}

In this present study, we seek to extend the PN approximations introduced by
\cite{Parisi2018,Yang2018} to explicitly calculate the gravitational waveform
of f-modes from highly eccentric BH--NS binaries.
The GWs from the orbital motion in this limit have already been
calculated and analyzed in \cite{Loutrel2019,Loutrel2020}.
Here, we adapt their effective fly-by framework to NS f-modes.
We work to leading order in the PN formalism,
decomposing the elliptic orbit as a Fourier series of harmonics of
the orbital period.
The f-modes are excited by this orbit and can be written exactly
as an integral of the Greens' function of the driving force,
which can then be solved using the harmonic Fourier series of the orbit.
However, such a series requires multiple terms to accurately
describe high eccentricities
and to capture the loudest harmonics.
We thus adapt the re-summation procedure of \cite{Loutrel2017a,Loutrel2019}
to integrate the series.
We find that the f-modes oscillate at their natural frequency
with an amplitude determined
by the stellar structure and the orbital parameters at closest approach,
further demonstrating the effective fly-by description of the
re-summation procedure.
Mathematically,
this appears through the evaluation of the series
at the terms closest to resonating with the dominant f-mode harmonic.
These coefficients determine the amplitude
of a simple sinusoidal tide.
This is our main result,
and we compare it to the numerical integration of the oscillations.
We find that our re-summations performs well for highly eccentric
and close pericenter passages,
reaching matches with the numerical integrations greater than 0.98
for eccentricities $e>0.8$ and semi-latus rectums $p<15 M$.

We also discuss how to account for subsequent mode excitations
with a radiation reaction model that determines
how the orbital parameters change between pericenter passages
and the timing between these passages.
By direct integration of the 2.5PN equations of motion, the orbital parameters of subsequent pericenter passages can be found via a recursion relation. For the times of pericenter passage, we perform a partial re-summation of the PN expanded orbital period. However, there is not enough information in the PN expanded expression to fully determine all of the parameters of the re-summation. We calibrated the free parameters of the re-summed time against numerical integrations of the PN equations of motion at 2.5PN order, and find that the calibrated model is capable of recovering the time of pericenter passage to at most $\sim1M$ at $p=20M$ and less than $1M$ at smaller semi-latus recta.

In Sec. \ref{sec:tides} we present the leading order dynamical tides
raised on the star and calculate its induced the quadrupole moment.
Expanding the orbit in a Fourier series,
we then write it as a function of time
and calculate its GW polarizations.
We review the effective fly-by framework in Sec. \ref{sec:efbs}
and apply it to the f-modes through contour integration,
and introduce our timing model to create a sequence of tides.
Sec. \ref{sec:accuracy} demonstrates the accuracy of our re-summed modes
and shows the match between them and numerical waveforms within the
sensitivities of LIGO and ET.
We conclude in Sec. \ref{sec:discussion}
by discussing the work needed to produce a complete eccentric BH--NS model.
All expressions are given in units $G = c = 1$.
\section{Dynamical Tides in the Post-Newtonian Formalism}
\label{sec:tides}
At closest approach, the black hole's tidal field strongly affects the neutron star, deforming it and exciting fundamental f-modes on its surface \cite{Poisson}.
In this section we review how this deformation perturbs the two-body point mass problem into one with an additional quadrupole moment induced by the raised tides.
The excitation of these modes can be expressed in terms of the binary's orbit,
which we then decompose into harmonics of the mean anomaly
allowing us to express the moment as a function of time
and thus obtain the GWs emitted by the perturbed star.

The dynamic tide raised on a star has been analyzed in detail by linearizing
the Newtonian fluid equations for a non-spinning body and finding the normal
modes of the fluid displacement
$\bxi(\bm{x}, t)$ \cite{{Turner1977, Press1977}}.
The fluid displacement obeys the equation 
\begin{equation}\label{eq:diffxi}
\frac{\partial^2 \bxi}{\partial t^2} + \mathcal{L} \bxi = \nabla \Utidal,
\end{equation}
where $\mathcal{L}$ is a self-adjoint operator representing the spatial derivatives of the perturbed Euler equations \cite{Poisson}
and $\Utidal$ is the tidal potential.
In this formalism, the displacement is decomposed into normal modes,
\begin{equation}
\bxi(\bm{x}, t) = \sum_{\lambda} Q_\lambda (t) \bxi_\lambda (\bm{x}),
\end{equation}
where the eigenfunctions are normalized by an inner product over the star's volume,
\begin{equation}
\int \bxi_{\lambda'}^* \cdot \bxi_\lambda \,\rho(\bm{x}) \dee^3 x = \delta_{\lambda' \lambda},
\end{equation}
with $\rho$ being the body's density and the asterisk denoting complex conjugation,
and are constrained by the eigenvalue equation
\begin{equation}
\mathcal{L} \bxi_\lambda - \omega^2_\lambda \bxi_\lambda = 0.
\end{equation}
By Fourier transformation, Eq. \eqref{eq:diffxi} yields a differential equation for the mode amplitude\footnote{Note that here we have defined the damping coefficient coupling to $\dot{Q}_{\lambda}$ to be $2\gamma$ instead of $\gamma$ as other references have. We do this to avoid certain factors of two that appear in the calculation to simplify expressions. The overall structure of the solutions does not change.} 
\begin{equation}\label{eq:diffQ}
\ddot{Q}_\lambda + 2\gamma_\lambda \dot{Q}_\lambda + \omega^2_\lambda Q_\lambda = U_\lambda.
\end{equation}
To account for dissipative mechanisms,
we have added the damping rate $\gamma_\lambda = 1/\tau_\lambda$ to account for a mode's dissipation within the fluid that occurs on a timescale $\tau_\lambda$.
The driving force is
\begin{equation}\label{eq:Ulambda}
U_\lambda (t) = \int \bxi_{\lambda}^* \cdot \nabla \Utidal \,\rho \dee^3 x.
\end{equation}
Eq. \eqref{eq:diffQ} can be solved as the superposition of a homogeneous
and an inhomogeneous solution.
Choosing the initial values at some time $t=0$,
the first solution is
\begin{multline}\label{eq:hom}
  Q^{\textrm{h}}_\lambda (t) = \exp(-\gamma_\lambda t)
  \Bigg[ Q_\lambda (0) \cos(\omega'_\lambda t) \\
  \left. + \frac{\dot{Q}_\lambda (0)  + \gamma_\lambda Q_\lambda(0) }{\omega'_\lambda} \sin(\omega'_\lambda t) \right],
\end{multline}
for the real frequency $\omega'_\lambda \equiv \sqrt{\omega^2_\lambda - \gamma_\lambda^2}$.
Using a Greens' function,
the inhomogeneous solution driven by $U_\lambda$ is
\begin{equation}\label{eq:Qinh}
  Q^{\textrm{inh}}_\lambda (t) = \frac{1}{\omega'_\lambda} \int_{-\infty}^{t}
  \exp[-\gamma_\lambda(t-t')] U_\lambda (t') \sin(\omega'_\lambda (t - t')) \,\dee t'.
\end{equation}

To begin integrating Eq \eqref{eq:Qinh}, we use spherical harmonics
to expand the tidal force of the black hole of mass $\MB$ \cite{Press1977},
\begin{equation}
\nabla \Utidal = \MB \sum_{lm} W_{lm} \frac{r^{l-1}}{R^{l+1}} \exp(-imV)
\left[ l \bm{e}_r + r \nabla \right] Y_{lm}(\Omega),
\end{equation}
where the $W_{lm}$ coefficients are defined in Ref. \cite{Press1977}'s Eq. 24, $R$ is the distance between bodies, and $(r, \Omega)$ describe a point in space centered at the neutron star, with $\bm{e}_r$ as the unit vector in the radial direction.
As will be explained in Sec. \ref{sub:harmdecomp}, $V$ is the orbit's true anomaly,
making it a function of time along with $R$.
Identifying the modes as the collection of integer indices $\{l, m\}$\footnote{
In wave theory, the indices are usually $\{n,l,m\}$,
but f-modes are not radial, so in our analysis the modes
are independent of $n$.
Furthermore, if the star is not spinning,
the eigenfunctions and frequencies are also independent of $m$ \cite{Yang2018, Kokkotas1999}.
}
, $m \in [ -l, l ]$, the eigenfunctions can be expanded similarly as 
\begin{equation}
\bxi_\lambda (\bm{x}) = \bm{\xi}_{lm} (r,\Omega) = \left[ \xi^R_{lm} (r) \bm{e}_r + \xi^S_{lm} (r) r \nabla \right] Y_{lm}(\Omega),
\end{equation}
where the (real) functions $\xi^R$ and $\xi^S$ can be found by considering
hydrostatic equilibrium and the boundary conditions on the star's surface \cite{Yang2018},
and the unperturbed star is assumed to be spherically symmetric.
Thanks to the orthogonality relations
\begin{align*}
\int Y_{lm} (\Omega) Y_{l'm'}^* (\Omega) \,\dee\Omega &= \delta_{ll'} \delta_{mm'}, \\
\int r\nabla Y_{lm}(\Omega) \cdot r\nabla Y_{l'm'}^*(\Omega) \,\dee\Omega &= l(l+1) \delta_{ll'} \delta_{mm'},
\end{align*}
we can express Eq. \eqref{eq:Ulambda} as
\begin{equation}\label{eq:Ulm}
U_{lm}(t) = \MB W_{lm} K_{lm} {\exp(-imV) \over R^{l+1}},
\end{equation}
where the overlap integral over the star's radius $r_*$ is
\begin{equation}
K_{lm} = l \int_0^{r_*} r^{l+1} \left[ \xi^R_{lm} (r) + (l+1) \xi^S_{lm} (r) \right] \,\rho \dee r.
\end{equation}
From these equations, the responses $Q_{lm}$ and frequencies $\omega_{lm}$ can be constrained \cite{Yang2018}.
Solving for the response of the star to the gravitational potential,
we can insert Eq. \eqref{eq:Ulm} into \eqref{eq:Qinh},
\begin{multline}
Q_{lm} (t) = \MB \frac{W_{lm} K_{lm}}{\omega'_{lm}} \\
\times \int_{-\infty}^t \exp[-\gamma_\lambda(t-t')] \frac{\exp(-imV(t'))}{R^{l+1}(t')} \sin(\omega'_{lm} (t - t')) \,\dee t'.
\end{multline}

The inhomogeneous solution is the response of the star to the driving potential $\Utidal$, which for a homogeneous star excites only fundamental f-modes \cite{Turner1977}.
In general,
not only are other perturbations, such as p- and g-modes, negligible
compared to these,
but the f-modes are primarily excited by the potential's quadrupolar moment $l = 2$.
Higher moments are suppressed by a factors of $R$,
and as $1/R \sim v^2$ (the square of the orbital velocity),
they constitute higher PN corrections.
Thus, this quadrupolar moment is the leading PN order description of the tide.

\subsection{Quadrupole Moment}
\label{sub:quadrupole}

To obtain the waveform of the neutron star's quadrupole tide, we consider the symmetric trace-free quadrupole moment \cite{Poisson},
\begin{equation}\label{eq:STFI}
\STFu{I}{jk}_{\textrm{tidal}} (t) = \frac{8\pi}{15} \sum_{m = -2}^2 I_{2m} (t) \Yc{jk}_{2m},
\end{equation}
where
\begin{equation}
I_{2m} (t) = \int \rho(t, \bm{x}) r^2 Y_{2m}^* (\Omega) \,\dee^3 x
\end{equation}
is the $\{l = 2, m\}$ multipole moment of the star,
with the asterisk indicating complex conjugation.
We have also used the spherical harmonic tensors $\STFu{\mathcal{Y}}{jk}_{2m}$ that are defined as (for $l = 2$)
\begin{equation}
\STFu{\mathcal{Y}}{jk}_{2m} = \frac{15}{8\pi} \int \STFu{n}{jk} Y_{2m}^* (\Omega) \,\dee\Omega,
\end{equation}
where $\STFu{n}{jk} = n^j n^k - {1\over3} \delta^{jk}$ for the unit vector $n^j$,
and $\STFu{\mathcal{Y}}{jk}_{l,-m} = (-1)^m \Yc{jk}_{lm}$. 
The use of symmetric trace-free tensors (denoted by the bracketed indices) serves us to expand the quadrupole moment in spherical harmonics, such as in Eq. \ref{eq:STFI}.

The utility of this formalism is more apparent when we find the tidal contribution
by perturbing the density of the neutron star as $\rho \to \rho + \delta\rho$, where in the Lagrangian description,
\begin{align}
\delta\rho =& -\nabla \cdot \left( \rho \bm{\xi} \right) \\
\begin{split}
=& - \sum_{lm} Q_{lm} (t) Y_{lm} (\Omega) \\
&\times
\left[ \frac{\partial}{\partial r} \left( \rho \xi_{lm}^R (r) \right) - \frac{\rho}{r} l(l+1) \xi_{lm}^S (r) \right],
\end{split}
\end{align}
recalling that $r^2 \nabla^2 Y_{lm} = -l(l+1) Y_{lm}$.
The multipole moment can be split into an integral over the unperturbed star plus its deformation.
As the star is spherically symmetric and stationary, its unperturbed density $\rho$ does not yield a time varying quadrupole moment, and in the absence of spin can be neglected.
We are then left with the moment of the density perturbation,
\begin{align}
\label{eq:I2m}
I_{2m} (t) &= \int \delta\rho~ r^2 Y_{2m}^* (\Omega) \,\dee^3 x \nonumber \\
&= - \varepsilon_{2m} Q_{2m}(t),
\end{align}
where
\begin{equation}
\varepsilon_{lm} = \int_0^{r_*}\left[ \frac{\partial}{\partial r} \left( \rho \xi_{lm}^R (r) \right)
- \frac{\rho}{r} l(l+1) \xi_{lm}^S (r) \right] r^4 \dee r.
\end{equation}

The quadrupole moment can now be found in terms of the tidal response.
We henceforth suppress the $l=2$ index for convenience.
As $W_{\pm1} = 0$, the relevant coefficients are
\begin{equation}
W_0 = \sqrt{{\pi \over 5}}, \quad W_{\pm2} = \sqrt{{3\pi \over 10}},
\end{equation}
with the respective tensors
\begin{align}
  \label{eq:y2m}
  \STFu{\mathcal{Y}}{jk}_0 &= \sqrt{\frac{5}{16\pi}} \diag(-1, -1, 2), \\
  \STFu{\mathcal{Y}}{jk}_2 &= \Yc{jk}_{-2} = \sqrt{\frac{15}{32\pi}}
  \begin{pmatrix}
    1 & -i & 0 \\
   -i & -1 & 0 \\
    0 &  0 & 0
  \end{pmatrix}.
\end{align}
We thus obtain the quadrupole moment
\begin{multline}\label{eq:I}
\STFu{I}{jk}_{\textrm{tidal}} (t) = -\frac{8\pi \MB}{15} \sum_{m = -2}^2 
\frac{\varepsilon_m W_m K_m}{\omega'_m} \Yc{jk}_m \\
\times \int_{-\infty}^t \exp[-\gamma_m (t-t')] \frac{\exp(-imV(t'))}{R^3(t')} \sin(\omega'_m (t - t')) \,\dee t'.
\end{multline}
The amplitude of the oscillation is thus EOS-dependent through
the constants $\varepsilon_m$ and $ K_m$.
\subsection{Harmonic Decomposition}
\label{sub:harmdecomp}

To find the quadrupole moment explicitly, we must integrate Eq. \eqref{eq:I}.
We investigate the time dependence of the distance between the black hole and the neutron star, $R$,
and the true anomaly, $V$.
In the Kepler problem of a binary with total mass $M = M_* + \MB$,
where $M_*$ is the neutron star's mass,
two other anomalies are known: the eccentric anomaly $u$ that can be related to the orbital phase as
\begin{equation}\label{eq:cosu}
\cos V = \frac{\cos u - e}{1 - e \cos u}
\end{equation}
and
\begin{equation}\label{eq:sinu}
\sin V = \frac{\sqrt{1-e^2} \sin u}{1 - e \cos u},
\end{equation}
where $e$ is the orbit's eccentricity;
and the mean anomaly
\begin{equation}\label{eq:ell}
\ell = u - e \sin u = n (t - \tp),
\end{equation}
where $n = \sqrt{M / a^3}$ is the orbital frequency for a binary of total mass $M$, semi-major axis $a$, and time of pericenter passage $\tp$.
Measured from the line of nodes, the orbit revolves by an angle $\phi = V + \varpi$
such that $\varpi$ is the longitude of pericenter, where the bodies are at closest approach.

These anomalies allow for useful representations of the orbit;
for example, Eq. \eqref{eq:cosu} allows for the equation of the orbit to be expressed as
\begin{equation}
R = \frac{a(1 - e^2)}{1 + e \cos V} = a (1 - e \cos u).
\end{equation}
However, there is no simple function $R(t)$ for non-circular orbits.
To use $V$, the differential equation
\begin{equation}
  \label{eq:orbit}
  \frac{\dee\phi}{\dee t} = \frac{\sqrt{Mp}}{R^2}
\end{equation}
has to be integrated for $V = \phi-\varpi$,
or to use $u$, Eq. \eqref{eq:ell} has to be solved numerically.
Here we have introduced the semi-latus rectum $p=a(1-e^2)$
of the orbital ellipse,
which is a useful alternative to the semi-major axis
for eccentric orbits.

To calculate the orbit as an explicit function of time,
we turn towards a more useful representation of the orbit:
its harmonic decomposition in $\ell$.
Consider expanding a function $\mathfrak{f}$ of the anomalies as a Fourier series
\begin{equation}
\mathfrak{f} = \sum_{k=-\infty}^{\infty} c_k \exp(ik\ell),
\end{equation}
where one can find the coefficients with
\begin{equation}\label{eq:fourier}
c_k = \frac{1}{2\pi} \int_{-\pi}^{\pi} \mathfrak{f} \exp(-ik\ell) \dee\ell.
\end{equation}
This expresses $\mathfrak{f}$ as a sum of epicycles.
With the integral representation of the Bessel function
\begin{equation}
J_k (x) = \frac{1}{2\pi} \int_{-\pi}^{\pi} \exp[i(ku - x \sin u)] \dee  u,
\end{equation}
the following can be found:
\begin{align}
  \label{eq:cosv}
  \cos V &= -e + \frac{2}{e} (1 - e^2) \sum_{k=1}^{\infty} J_k (ke) \cos(k\ell), \\
  \label{eq:sinv}
  \sin V &= 2 \sqrt{1 - e^2} \sum_{k=1}^{\infty} J'_k (ke) \sin(k\ell), \\
  \frac{a}{R} &= 1 + 2 \sum_{k=1}^{\infty} J_k (ke) \cos(k\ell),
\end{align}
where $J'_k (x) = \frac{\dee}{\dee x} J_k (x)$.
One can thus express the orbit explicitly as a function of time, albeit in terms of infinite sums.

\subsubsection{Hansen Coefficients}
\label{sub:hansen}

The orbital functions $\cos V$, $\sin V$, and $a/r$ were expressed as Fourier series of the orbit's epicycles in the previous section.
However, in integrating Eq. \eqref{eq:I}, we have the task of Fourier expanding the terms
$$\frac{1}{R^3}, \quad \frac{\exp(\pm i2V)}{R^3}.$$
We follow \cite{Yang2018}, considering the general expansion
\begin{equation}\label{eq:hansen}
\left( \frac{R}{a} \right)^q \exp(imV) = \sum_{k = -\infty}^{\infty} X^{q,m}_k (e) \exp(ik\ell).
\end{equation}
We see that the quadrupole moment is composed of the coefficients $(q,m) = (-3,\pm2)$ and $(-3,0)$.
The $X^{q,m}_k$ are known as the Hansen coefficients \cite{tisserand1889}.

With these coefficients, we can now integrate Eq. \eqref{eq:I} to find the star's response to the driving potential,
\begin{multline}
\int_{-\infty}^t \exp[-\gamma_m (t-t')] \frac{\exp(-imV)}{R^3} \sin(\omega'_m (t - t')) \dee t' =\\
\frac{\omega'_m}{a^3} \sum_{k=-\infty}^{\infty} X^{-3, -m}_k \frac{\exp(ik\ell)}{{\omega}_m^2 - (kn)^2 + 2ikn\gamma_m},
\end{multline}
yielding the response
\begin{equation}\label{eq:tide}
  Q_{m} (t) = \frac{\MB}{a^3}  W_{m} K_{m} \sum_{k=-\infty}^{\infty}  \frac{X^{-3, -m}_k\exp(ik\ell)}{\omega_m^2 - (kn)^2 + 2ikn\gamma_m}.
\end{equation}
Here we have assumed that the modes were quiet in the past, such that this is the response for a single passage.
Tides raised by a past passage can be added coherently
using the homogeneous solution (Eq. \eqref{eq:hom}).
The initial conditions for a tide can be evaluated knowing the time
of each pericenter passage.

It is expected that the series \eqref{eq:hansen} will converge more slowly
as the eccentricity approaches 1.
As one can also see from their definition in Eq. \eqref{eq:fourier},
the Hansen coefficients can become difficult to evaluate numerically for
large $k$ as the integrand becomes highly oscillatory.
These two facts complicate our analysis.
As $\omega_m$ lies in the kHz frequencies for NSs,
one can see in Eq. \eqref{eq:tide} that the dominating terms of the series
are of large index $k$ such that the f-modes are close to resonance
with the orbital frequency.
The coefficients of interest cannot be found in simple analytic form,
unless expanded in another series of their own \cite{Sadov2006,Sadov2008}.

\subsection{Gravitational Waves From f-Modes}
\label{sub:gws}

In Sec. \ref{sub:quadrupole} and \ref{sub:harmdecomp}, we calculated
the f-mode excitation and induced quadrupole moment of a star.
Eq. \eqref{eq:I} yields this tidal response as a function of time.
It is then straightforward to find the leading order gravitational waves
from these f-modes with the quadrupole formula for the metric perturbation
far from the source,
\begin{equation}
\label{eq:h}
h_{jk} = \frac{2}{d_{\rm L}} \ddot{I}_{jk},
\end{equation}
where $I_{jk}$ is the quadrupole moment of the sources, and $d_{\rm L}$ is the luminosity distance to the origin of our coordinate system.
Besides the moment from the motion of the binary,
$I_{jk}$ also includes the tidal response from the neutron star,
$I_{jk}^{\rm{mode}}$,
so the total quadrupole moment is the orbit plus the tide \cite{Will1983}.
However, we are interested primarily in the signal from the tidal excitations,
and so the orbit's radiation is excluded in this study.
The gravitational waves from eccentric binaries at leading PN order
have been studied in \cite{Loutrel2019}.

To analyze the observable effects of the metric perturbation,
we project it into the transverse-traceless gauge.
Setting our coordinate system on the orbital plane and centered at its center of mass,
the unit vector pointing to an observer is
\begin{equation}
\label{eq:d}
\bm{\hat{d}} = \left[ \sin\iota \cos\psi, \sin\iota \sin\psi, \cos\iota \right],
\end{equation}
where $\iota$ is the inclination angle to the binary's orbital angular momentum
and $\psi$ is an arbitrary polarization angle.
To describe the space orthogonal to $\bm{\hat{d}}$, we also define
\begin{align}
  \label{eq:Theta}
  \bm{\Theta} &= \left[ \cos\iota \cos\psi, \cos\iota \sin\psi, -\sin\iota \right], \\
  \label{eq:Phi}
  \bm{\Phi} &= \left[ -\sin\psi, \cos\psi, 0 \right].
\end{align}
Although the position of the neutron star (the source of $h_{jk}$)
is not at the center of our coordinates,
the separation between them is negligible compared to the distance
expected from astrophysical sources.
Thus, we can approximate the line-of-sight from the star as
the line-of-sight from the center of mass to leading order in $1/d_{\rm L}$,
as we show in Appendix \ref{app:los}.

\begin{figure}[t]
  \includegraphics[width=0.45\textwidth]{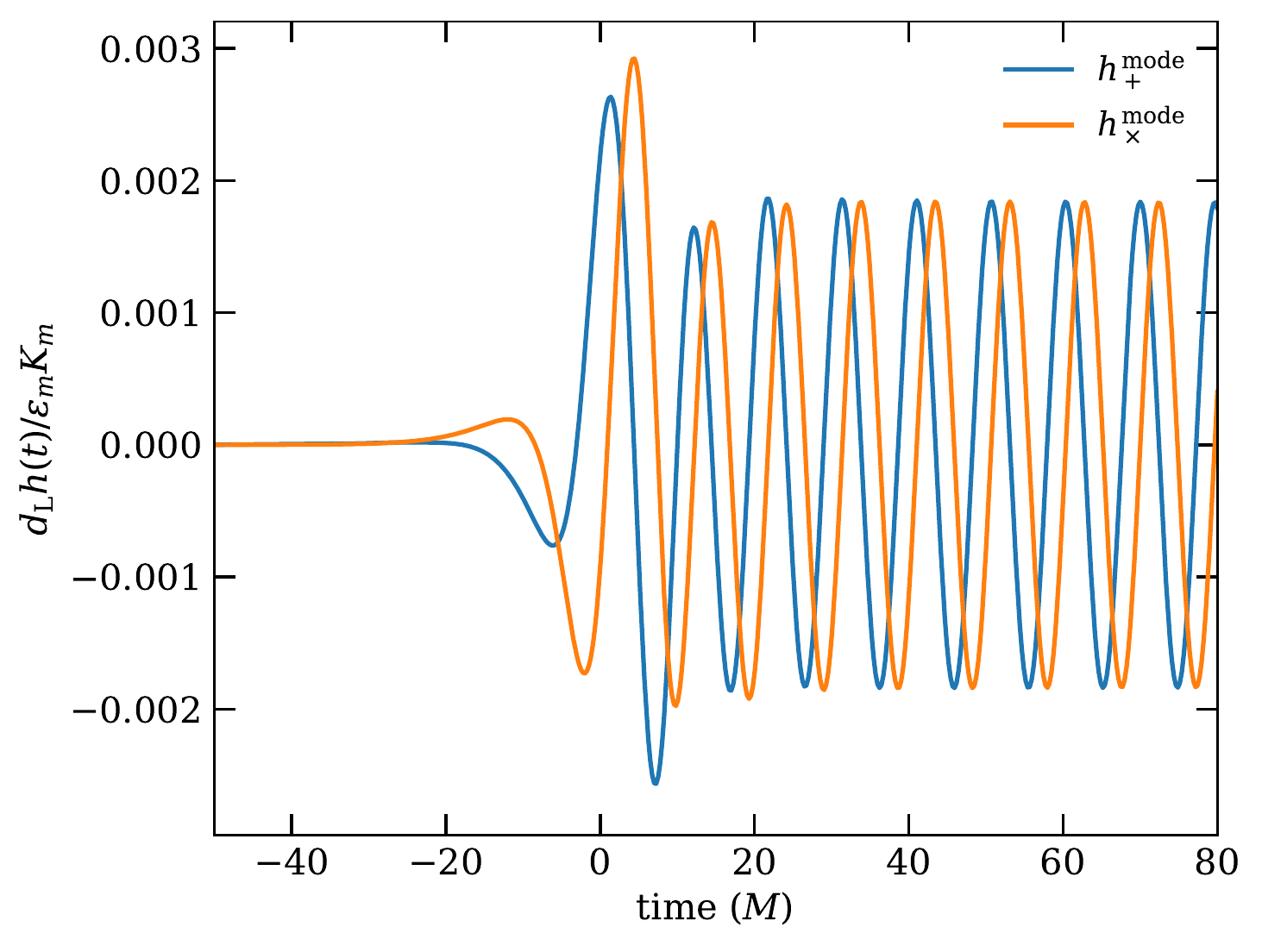}
  \caption{Plus and cross polarizations of a tidal excitation with
    $\iota = \psi = 0$ (face-on binary) scaled by the luminosity distance,
    where the BH and NS are closest to each other at $t=0$.
    The NS has an f-mode frequency $f = \omega/2\pi = 1.865$ kHz
    and dampening timescale $\tau = 0.230$ s,
    which is longer than the orbital period
    $T_{\rm orb} = 2399 ~\text{M} \approx 0.133$ s
    ($e = 0.9, p=10 M$).}
  \label{fig:h}.
\end{figure}

The plus and cross GW polarizations due to the tide can then be found to be \cite{Poisson}
\begin{align}
h_+ &= \frac{1}{2} \left( \Theta^j \Theta^k - \Phi^j \Phi^k \right) h_{jk}, \\
h_\times &= \frac{1}{2} \left( \Theta^j \Phi^k + \Phi^j \Theta^k \right) h_{jk}.
\end{align}
Inserting Eq. \eqref{eq:STFI} into Eq. \eqref{eq:h} yields the polarizations
\begin{align}
  h_+^{\rm mode} =& \frac{1}{3d_{\rm L}} \sqrt{\frac{2\pi}{5}} (1 + \cos^2\iota)
          \left[\ddot{I}_2 \exp(i2\psi) + \ddot{I}_{-2} \exp(-i2\psi)\right] \nonumber \\
         &+ \frac{2}{d_{\rm L}}\sqrt{\frac{\pi}{5}} \sin^2\iota \ddot{I}_0 , \\
  h_\times^{\rm mode} =& \frac{i2}{3d_{\rm L}} \sqrt{\frac{2\pi}{5}} \cos\iota
          \left[ \ddot{I}_2 \exp\left(i2\psi\right)
           - \ddot{I}_{-2} \exp\left(-i2\psi\right) \right].
\end{align}
We show examples of these polarizations in Fig. \ref{fig:h}
for an eccentric orbit ($e = 0.9, p=10 M$) containing a 1.273 M$_\odot$ NS with the
SLy4 equation of state (see Table XI in \cite{Chirenti2015})
and a 10 M$_\odot$ BH.
The $\ddot{I}_m$ are calculated from Eq. \eqref{eq:I2m} by numerically
integrating $Q_m (t)$ as described in Sec. \ref{sec:accuracy},
where we compare the polarizations to the analytic expressions
developed in Sec. \ref{sec:efbs}.

\section{$f$-Modes in Effective Fly-Bys}
\label{sec:efbs}

While in Section \ref{sec:tides} we provided an analytic description
of the tides raised on a star in an arbitrary binary,
we now focus on eccentric binaries.
We've mentioned that the series in orbital harmonics Eq. \eqref{eq:hansen}
converges more slowly as $e \to 1$ due to the deviation of the orbit from circular,
requiring more and more epicycles to accurately represent the orbit.

Alternative methods such as post-circular expansions in $e$ have sought
to extend the range of circular models.
Recently, a \emph{re-summation} method that instead expands
the harmonics in large $e$
has been investigated for hereditary fluxes \cite{Loutrel2017a}
and resulted in waveforms for eccentric binaries
to leading order \cite{Loutrel2019}.
These waveforms replace Eqs. \eqref{eq:cosv}-\eqref{eq:sinv}
with their asymptotic expansions about $k\gg1$ and re-sum them by replacing
their summation with an integral over $k$.
This method yields accurate representations of eccentric orbits,
eliminating the need to sum over many terms.
In this representation, the oscillatory orbit
is replaced by a post-parabolic orbit,
thus more accurately describing an effective fly-by.
We now apply this re-summation method to f-modes as follows.

\subsection{Re-summation of $f$-Modes}
\label{sub:resummation}

\begin{figure}[t]
  \includegraphics[width=0.5\textwidth]{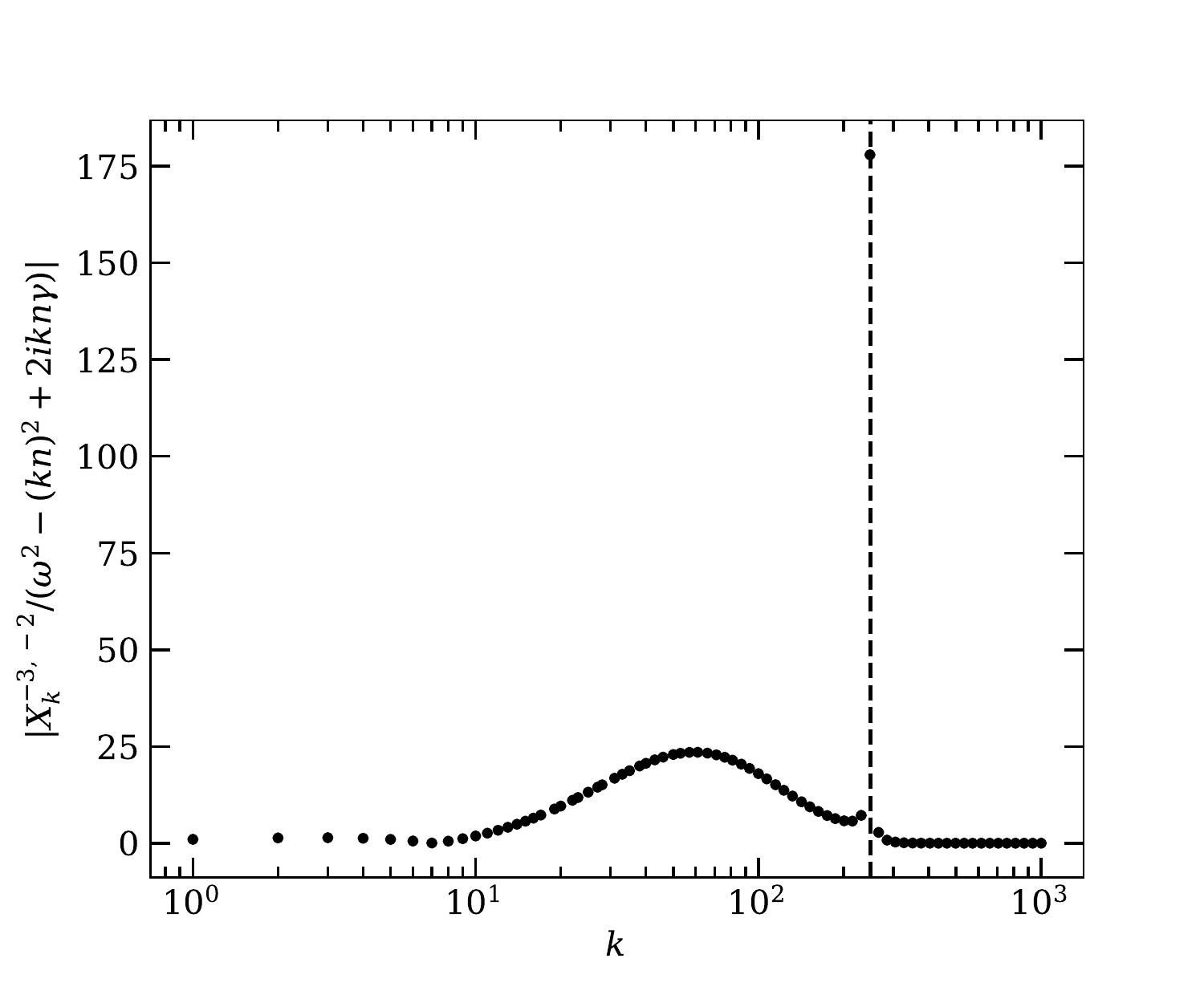}
  \caption{The absolute value of the Fourier coefficients of $Q_2$
    for the same system in Fig. \ref{fig:h}.
    The dashed line marks where $k = \omega/n \approx 248.4$.
    The bump in the tens of $k$ corresponds to the lower harmonics
    present in the series.}
  \label{fig:resonance}
\end{figure}

Consider the tide we found in Eq. \eqref{eq:tide}.
As $k$ ranges through all integers, there will be values at which
it will come close to resonating with the f-mode,
especially as this frequency is expected to surpass the orbit and the damping timescale.
Therefore the dominant term in the series is not at low values of $k$,
but rather at high values where the resonance dominates,
as shown in Fig. \ref{fig:resonance}. With the dominant terms being at large $k$,
we can re-sum our series with an integral as
\begin{equation}\label{eq:resummation}
  \sum_{k=-\infty}^\infty \to \int_{-\infty}^\infty \dee k,
\end{equation}
since all epicycles contribute to the sum in the high eccentricity limit.

\begin{figure*}[t]
  \includegraphics[width=0.49\textwidth]{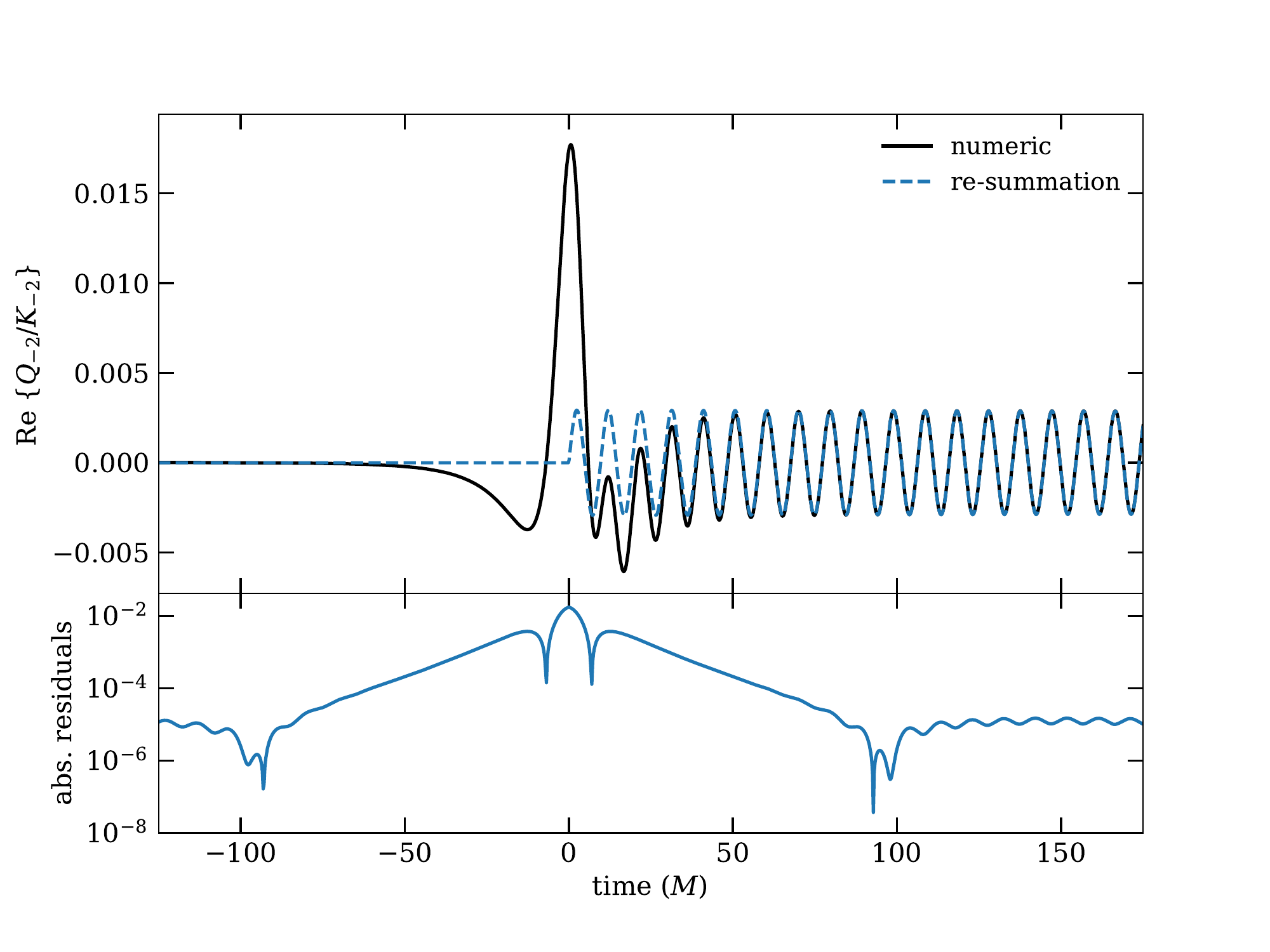}
  \includegraphics[width=0.49\textwidth]{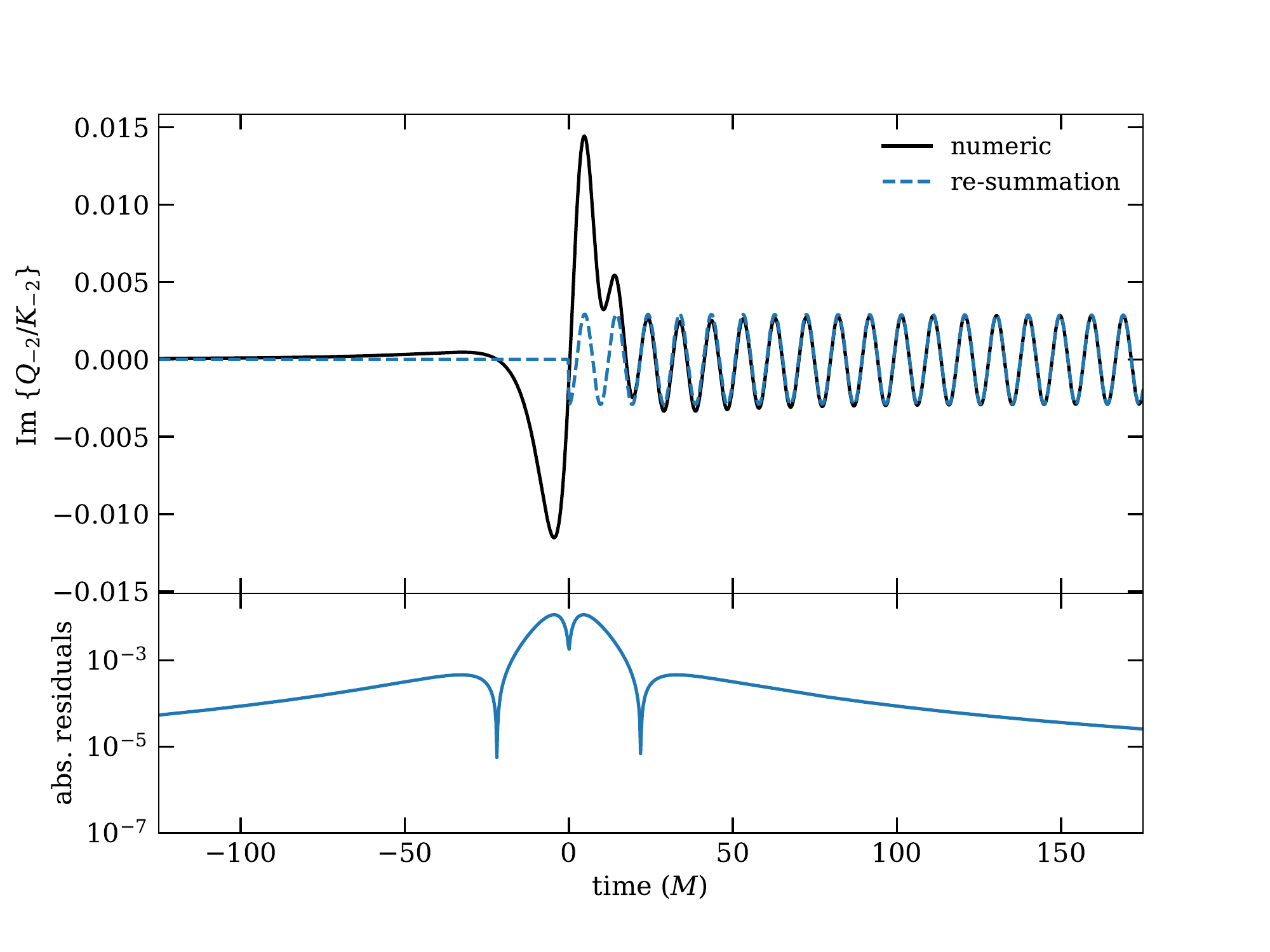}
  \caption{The real (left) and imaginary (right) amplitudes of the $m=-2$ mode
    for the same system in Fig. \ref{fig:h}.
    The numerically integrated modes are in solid black lines,
    the re-summed modes are the blue dashed lines.
    The bottom panels show the absolute value of the residuals between them,
    which reach a steady state away from the closest passage.
  }
  \label{fig:resum}
\end{figure*}

Applying this to the tide in Eq. \eqref{eq:tide},
we now have the task of integrating
\begin{equation}
  \int_{-\infty}^{\infty} \frac{ X^{-3,-m}_k \exp(ik\ell)}{\omega^2 - (kn)^2 + 2ikn\gamma} \dee k.
\end{equation}
Note that the integrand has poles at
\begin{equation}
  \label{eq:kpm}
  k_\pm \equiv \frac{1}{n} ( i\gamma \pm \omega'),
\end{equation}
which lie on the positive imaginary plane.
We can then make the integral as part of a contour integral over this plane,
and is thus accompanied by an integral over a semi-circle on it,
with our integral as its border on the real axis.
The integral over the semi-circle vanishes as
$\lim_{k\to +i\infty} \exp(ikl) \to 0$ for $\ell > 0$;
for $\ell < 0$, we would place the semi-circle on the negative
imaginary plane where the absence of poles would
make the contour integral vanish by Cauchy's theorem.
Only the integral on the real axis is left,
and we can cast the contour integral as a sum of residues:
\begin{multline}
  \int_{-\infty}^{\infty} \frac{ X^{-3,-m}_k \exp(ik\ell)}{\omega^2 - (kn)^2 + 2ikn\gamma} \dee k = \\
  2\pi i \sum_{k_\pm}
  \Res\left[ \frac{X^{-3,-m}_k \exp(ik\ell)}{\omega^2 - (kn)^2 + 2ikn\gamma}, k_\pm \right].
\end{multline}
Note that this applies only for $\ell > 0$, that is, after pericenter passage.
In the high eccentricity limit,
the tide is raised at closest approach,
with the black hole's potential ringing the neutron star like a bell.
Evaluating the residues
\begin{multline}
  \Res\left[ \frac{X^{-3,-m}_k \exp(ik\ell)}{\omega^2 - (kn)^2 + 2ikn\gamma}, k_\pm \right] \\
    =\mp \frac{X^{-3,-m}_{k_\pm} \exp\left[ -\frac{\ell}{n} (\gamma \mp i\omega') \right]}{2n\omega'},
\end{multline}
we obtain the tide
\begin{multline}
  \label{eq:resumQ}
  Q_m(t) = \MB \left( \frac{1-e^2}{p} \right)^3  W_{m} K_{m}\frac{\pi i}{n\omega'} \exp(-\gamma\Delta t) \\
  \times \left(
    X^{-3,-m}_{k_-} \exp\left( -i\omega' \Delta t \right) -
    X^{-3,-m}_{k_+} \exp\left( i\omega' \Delta t \right) \right),
\end{multline}
where $\Delta t \equiv t - \tp$. We refer to this as the re-summed f-mode.

Naively, it appears the re-summed mode in Eq.~\eqref{eq:resumQ} vanishes in the parabolic limit ($e\rightarrow1$). However, in order to accurately study this limit, one must consider the eccentricity dependence of the Hansen coefficients. While we do not have an analytic expression for the Hansen coefficients,
their definition shows divergence as $e$ approaches unity.
The divergence can be shown explicitly in $X^{-3,0}_{0}$, which can be calculated in closed form,
\begin{equation}
  X^{-3,0}_0 = (1-e^2)^{-3/2}.
\end{equation}
The remaining coefficients $X^{-3,m}_{k \neq 0}$  do not admit exact closed-form expressions, and we are forced to numerically evaluate them. Generically, these coefficients are enhanced, and appear to diverge, as the eccentricity approaches unity. 
We have numerically shown that multiplying $X^{-3,-m}_k$ by $(1-e^{2})^{3/2}$ for $m=0,\pm 2$
removes this divergence for $k$ up to 10,000. Thus, we postulate that we can write $X^{-3,m}_{k}(e) = \hat{X}^{-3,m}_{k}(e)/(1-e^{2})^{3/2}$, where $\hat{X}^{-3,m}_{k}(e)$ is regular for all values of $e$.
Further, one must remember that $n \propto (1-e^2)^{3/2}$, and thus, it is now clear that the product
$(1-e^2)^3 X^{-3,-m}_k/n$ in the amplitude of the re-summed mode neither diverges nor vanishes in the
limit of high eccentricity.
The star thus responds as a damped harmonic oscillator and is well behaved in the parabolic limit.

An interesting effect of the re-summation is the promotion of $k$
from integer to complex number, which also causes the Hansen coefficients
to become complex-valued.
The symmetry $X^{-3,m}_{k_+} = X^{-3,-m}_{k_-}$ appears,
which aids in evaluating these coefficients.
Nonetheless, the lack of an analytic form for $X^{-3,m}_k$ keeps us
from obtaining a completely analytic model,
and for now we continue numerically calculating the coefficients
using the Python package SciPy's \verb|quad| module.

Fig. \ref{fig:resum} shows how well the re-summation Eq. \eqref{eq:resumQ}
approximates the numerically integrated $Q_m$ for $m=-2$. For the numerical integration, we start the binary at apocenter half an orbital period $T_{\rm orb}$ away from closest approach and assume that there is no pre-existing mode driving on the star, specifically $Q_m(-T_{\rm orb}/2) = 0 = \dot{Q}_{m}(-T_{\rm orb}/2)$.
We then allow the binary to evolve through the next pericenter
and up to 100 f-mode cycles after it.
Away from pericenter passage, the residual between the numerical and re-summed modes are on the order of $10^{-5}$
and is phase-accurate.
Our approximation, however, does miss the large amplitude excitation
at closest passage, which is caused by the low $k$ values in the sum in Eq.~\eqref{eq:tide}.
Similar results are obtained for the other $m$ modes.

\begin{figure}[t]
  \includegraphics[width=0.5\textwidth]{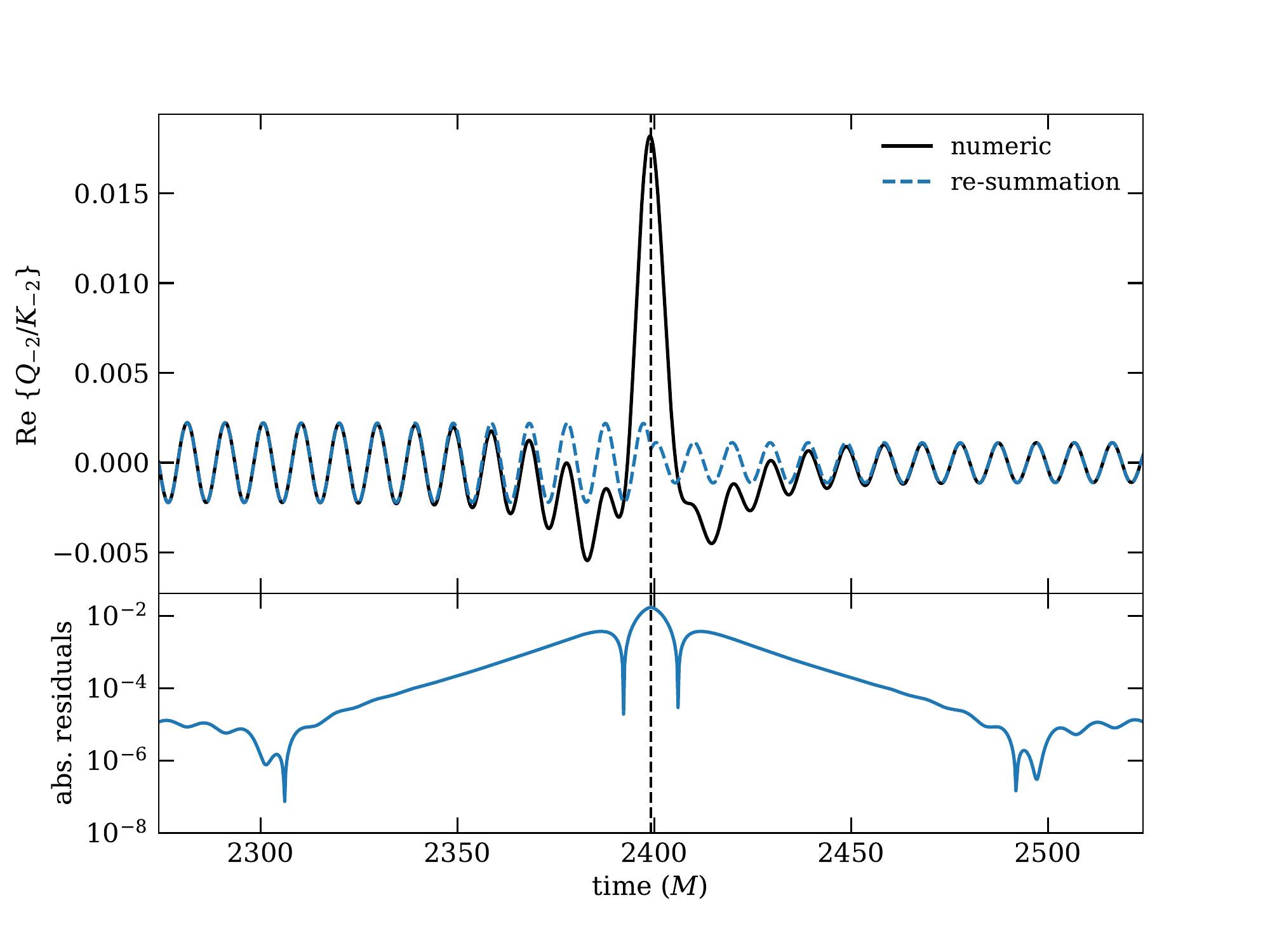}
  \caption{The real amplitude of the $m=-2$ mode across the first
    pericenter passage after the one in Fig. \ref{fig:resum}.
    While the large amplitude excitation is not captured by the re-summation,
    it remains accurate after it,
    the value of the residuals remaining as small.
    Similar results are obtained for the imaginary amplitude.}
  \label{fig:stitch}
\end{figure}

If no f-modes are present in the star before a passage,
only the inhomogeneous solution to the wave equation in Eq.~\eqref{eq:diffQ} is excited.
On subsequent passages, however, past excitations must be taken into account
if they haven't been dissipated away due to GW emission or viscosity in the star.
At the close separations that the star and BH must be for a significant tide
to be excited, the time between passages may not be enough
for this dissipation to have dampened the modes away.
Therefore to construct a sequence of tidal excitations we must add the modes
from previous passages.
This can be done by using the homogeneous solution \eqref{eq:hom},
evaluating the initial conditions $Q(t=t_{\mathrm{p}, N+1})$
and $\dot{Q}(t=t_{\mathrm{p}, N+1})$ at the $N+1$ pericenter passage,
and adding it to the first excited tide described by Eq. \eqref{eq:resumQ}.
This requires knowing each $t_{\mathrm{p}, N}$ and the value of $n$ at that time.
We demonstrate the results of this procedure in Fig. \ref{fig:stitch},
where $n$ is a constant without radiation or tidal back-reaction.
The expectation is that with a model detailing the evolution of $n$,
one can evaluate the initial conditions in the same way
and obtain a sequence of tides that remain as accurate.

\subsection{Timing Model}
\label{sub:timing}

The dynamical tide we obtained in Sec. \ref{sub:resummation}
responds to the BH's tidal potential,
its amplitude depending on the value of $n$ at $t_{\mathrm{p}, N+1}$
(see Eq. \eqref{eq:kpm}).
Were this a simple Keplerian orbit,
this would be a known constant, as in Fig \ref{fig:stitch}.
However, the orbit is radiating away energy and angular momentum,
necessitating a radiation reaction model that informs the evolution
of the orbit.
We now detail our model.

To leading order, the radiation reaction of an orbit appears at 2.5PN.
The orbit loses energy and angular momentum to gravitational radiation,
changing its Newtonian description over time.
We parametrize the osculating orbit with $p$ and $e$,
which then defines $n = \sqrt{M(1-e^2)^3 / p^3}$
and the energy and angular momentum as
$E = -M^2\eta (1-e^2)/2p$
and $L = \eta \sqrt{M^3 p}$, respectively,
where $\eta \equiv \MB M_* / M^2$ is the symmetric mass ratio.
In the harmonic gauge \cite{Poisson},
the orbital evolution is described by
\allowdisplaybreaks[4]
\begin{widetext}
\begin{align}
  \label{eq:dpdt}
  \frac{\dee p}{\dee t} &=
  -\frac{16}{5} \eta \left(\frac{M}{p}\right)^3 (1 + e\cos(V))^3
  (4+ e^2 + 5e\cos(V)),\\
  \label{eq:dedt}
  \frac{\dee e}{\dee t} &=
  -\frac{2}{15} \frac{\eta}{M} \left(\frac{M}{p}\right)^4 (1 + e\cos(V))^3
  \left[ (96+109e^2)\cos(V) + e(104+6e^2+2(56+9e^2)\cos(2V) + 35e\cos(3V)) \right],\\
  \label{eq:dwdt}
  \frac{\dee \varpi}{\dee t} &=
  -\frac{8}{15} \frac{\eta}{eM} \left(\frac{M}{p}\right)^4 (1 + e\cos(V))^3
  \left[6(4+e^2) + e(56+9e^2)\cos(V) + 35e^2\cos(V)\right] \sin(V),
\end{align}
\end{widetext}
which also changes the evolution of $V$ in Eq. \eqref{eq:orbit} to
\begin{equation}
  \label{eq:dvdt}
  \frac{\dee V}{\dee t} = \frac{\sqrt{Mp}}{R^2} - \frac{\dee \varpi}{\dee t}.
\end{equation}
Averaging these equations over an orbit
yields the adiabatic approximation, first calculated by Peters~\cite{Peters1964},
\allowdisplaybreaks[4]
\begin{align}
  \left< \frac{\dee p}{\dee t} \right> &=
  -\frac{64}{5} \eta \left(\frac{M}{p}\right)^3 (1-e^2)^{3/2} \left( 1+\frac{7}{8}e^2 \right),\\
  \left< \frac{\dee e}{\dee t} \right> &=
  -\frac{304}{15} \frac{e\eta}{M} \left(\frac{M}{p}\right)^4 (1-e^2)^{3/2} \left( 1+\frac{121}{304}e^2 \right).
\end{align}
Further, the secular change in $\varpi$ vanishes.
However, this approximation can lose accuracy for small $p$ and large $e$,
affecting the timing model.

The accuracy of the timing model is key to calculating the time of
pericenter passage and their Hansen coefficients.
One can obtain the orbital parameters $\lbrace p_{N+1}, e_{N+1} \rbrace$ at the
$N+1$ passage by considering that they change primarily at pericenter,
where almost all of the radiation occurs.
A model for these can then be obtained from the orbit-averaged evolution
\cite{Loutrel2017b,Loutrel2019}:
\begin{equation}
  \label{eq:p_model}
  p_{N+1} = p_N \left[1 - \frac{128\pi}{5} \eta \left(\frac{M}{p_N}\right)^{5/2}
    \left(1 + \frac{7}{8}e_N^2\right)\right],
\end{equation}
\begin{equation}
  \label{eq:e_model}
  e_{N+1} = e_N \left[1 - \frac{608\pi}{15} \eta \left(\frac{M}{p_N}\right)^{5/2}
    \left(1 + \frac{121}{304}e_N^2\right)\right].
\end{equation}
This yields a sequence of orbital parameters at each pericenter passage, which determine the amplitude of the f-modes.

To obtain the time at which these passages occur,
we consider that the evolution of the orbit is determined by the solution to the osculating equations, that is, being an initial value problem, the time from pericenter passage $t_{\mathrm{p},N}$ to the next depends on $p_N$ and $e_N$.
We infer that a function of the form
\begin{equation}\label{eq:t_model}
	t_{\mathrm{p},N+1}-t_{\mathrm{p},N} = \frac{2\pi}{M^{1/2}} \left(
	\frac{p_N+\eta \left(\frac{M}{p_{N}}\right)^{5/2}A}{\epsilon_N + \eta\left(\frac{M}{p_{N}}\right)^{5/2}B} \right)^{3/2},
\end{equation}
where $\epsilon \equiv 1-e^2$, describes the effects of radiation reaction, as suggested by the period of a Newtonian orbit.
The functions $A$ and $B$ are as of yet unknown, but from the fact that, for fixed values of $p_N$, $t_{\mathrm{p},N+1}-t_{\mathrm{p},N}$ increases monotonically with $e$, we propose that they can be well described by the polynomials
\begin{align}
	A &= a_0 + a_1 \epsilon_N, \\
	B &= b_0 + b_1 \epsilon_N + b_2 \epsilon_N^2,
\end{align}
where the $a$ and $b$ coefficients may depend on $\eta$ and $p_N$.
We then numerically generate values of $t_{\mathrm{p},N+1}-t_{\mathrm{p},N}$ across values of $p_N$, $e_N$, and $\eta$ and fit the coefficients to be, to double precision,
\begin{subequations}\label{eq:coeffs}
\begin{align}
	a_0 &= a_{0,1} + a_{0,2} \left(\frac{\eta}{1/4}\right)^{a_{0,3}} \left(\frac{p}{10M}\right)^{a_{0,4}},\\
	a_1 &= -16.823395797589278,\\
	b_0 &= 170\pi/3 = 178.0235837034216,\\
	b_1 &= -139.3766232947201,\\
	b_2 &= -1.088578314814299,
\end{align}
\end{subequations}
where
\begin{subequations}\label{eq:a0coeffs}
\begin{align}
	a_{0,1} &= 14.1774066465967,\\
	a_{0,2} &= -0.236903393660227,\\
	a_{0,3} &= 0.962439591179757,\\
	a_{0,4} &= -2.415912280582671.
\end{align}
\end{subequations}
We describe our procedure for finding these values in Appendix \ref{app:timing},
where we also demonstrate the excellent agreement between our model and the values of $t_{\mathrm{p},N+1}-t_{\mathrm{p},N}$.

This timing model yields highly accurate results on orbital timescales
when comparing the orbital parameters to the values found
with numerical integration.
This thus builds a sequence of pericenter passages,
from which we can evaluate the tides raised on the star.
We explore the accuracy of this model for the tides in the next section.

\section{Accuracy of the Re-summed Modes}
\label{sec:accuracy}

With the re-summation of the f-modes presented in Sec. \ref{sub:resummation}
and timing model from Sec. \ref{sub:timing},
we now have an analytic model for the leading-order GWs of these oscillations,
save for the Hansen coefficients, which we calculate numerically in this study
for the reasons stated in Sec \ref{sub:hansen}.

In this section we compare the re-summation to numerical integration of the
orbit and the f-modes.
The integration of the latter is performed by solving Eq. \eqref{eq:diffQ}
with the driving force Eq. \eqref{eq:Ulm} using the 4(5)th order
Runge-Kutta method from SciPy's \verb|solve_ivp|.
Simultaneously, the orbit is integrated with Eq. \eqref{eq:dvdt}
while evolving $p$, $e$, and $\varpi$ with Eqs. \ref{eq:dpdt} - \ref{eq:dwdt}.
This provides a 2.5PN numerical model of the orbit
that excludes tidal effects.
While it has been shown that the tide also causes the orbit to decay
faster than from radiation reaction alone \cite{Yang2018,Vick2019},
it is generally sub-dominant to the gravitational radiation.
The combined effects of radiation reaction and the tide will be
investigated in a future work.

\subsection{Comparison to Numerics}
\label{sub:comp}

We start our numerical integrations
at apocenter $t_0 = -T_{\rm orb,0}/2$, where $T_{\rm orb,0}$ is the period of the orbit at $t_0$.
As the orbit progresses, $e$ and $p$ decrease due to radiation reaction.
We integrate the orbit until either $t=T_{\rm orb,1}$ or $p < 2M(3+e)$,
the latter corresponding to the last stable orbit for Schwarzschild geodesics.
With $V$, $p$, $e$, $Q_m$, and $\dot{Q}_m$ calculated,
we can evaluate Eq. \eqref{eq:diffQ} to find $\ddot{I}_m$
and their GWs.

We set the initial conditions $Q_m(t_0) = \dot{Q}_m(t_0) = 0$
for the f-modes.
Physically, these initial conditions are only valid when the binary is infinitely far apart on an unbound orbit, such that the tidal force can be neglected. If starting the evolution at a finite separation as we are doing here, one would expect the modes
to not be quiet for any $e_{0} < 1$. Essentially, the binary will have to have evolved from a dynamical capture to an orbit with the specific $e_{0}$, and thus modes would have been excited from previous pericenter passages. However, if we simply desire to study the accuracy of the re-summed mode, it suffices to compare to numerical evolutions with the initial conditions $Q_{m}(t_{0}) = 0 = \dot{Q}_{m}(t_{0})$. It is worth noting that using such data at a finite orbital separation results in a low-amplitude mode being activated before the initial pericenter passage, especially when $e_{0} < 0.9$. The mode is unphysical, and is a result of not applying suitable initial data. One could use a reduction scheme, similar to eccentricity reduction in numerical relativity, to remove the mode, but we have found that this is not necessary. Instead, one can simply take this into account using the procedure described in Sec.~\ref{sub:resummation}.

\begin{figure}[t]
  \includegraphics[width=0.5\textwidth]{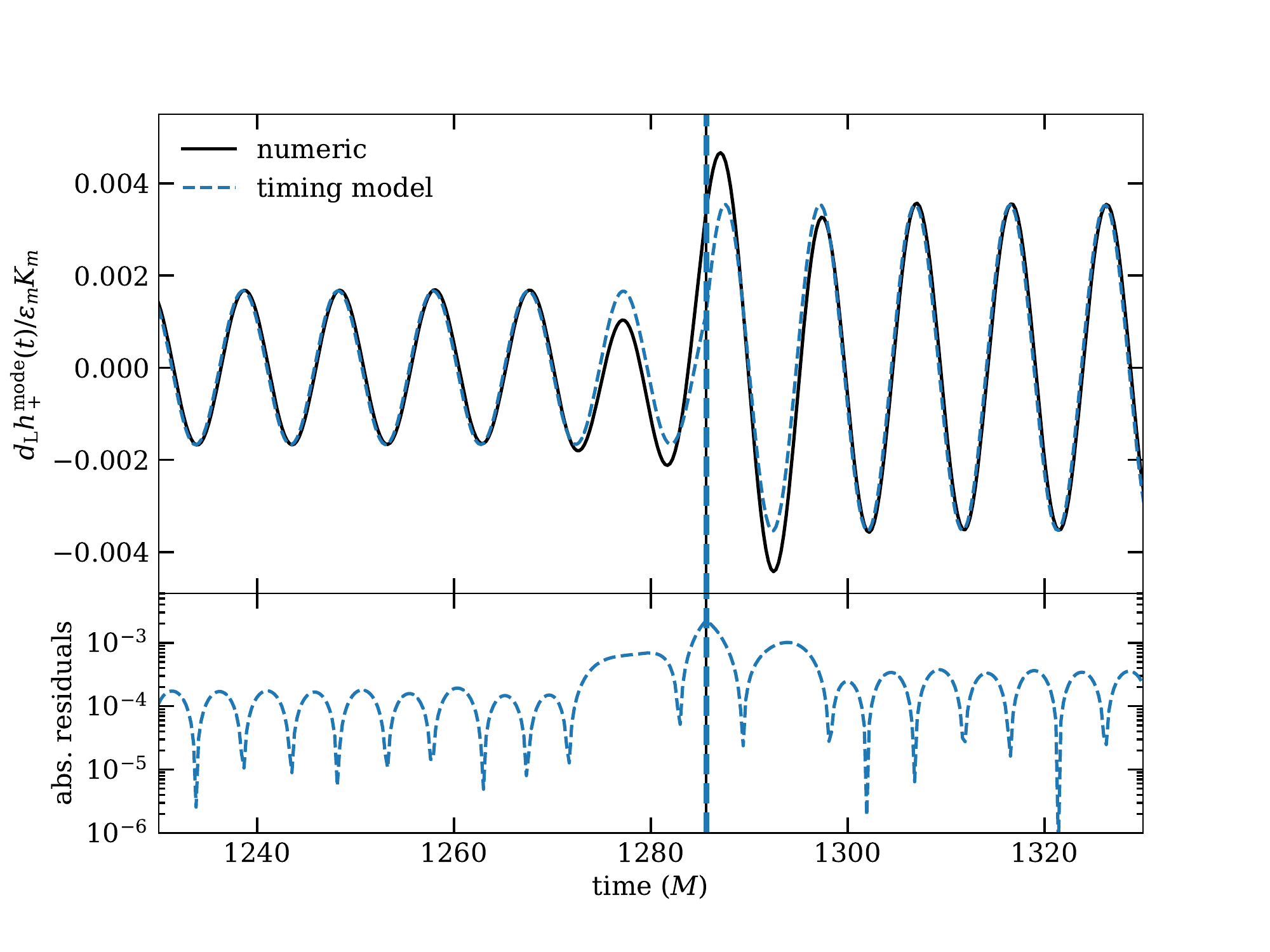}
  \caption{The plus polarization around the second passage of an orbit with $e_0 = 0.9$, $p_0 = 10 M$,
    modeled with the numeric integration (solid black),
    the timing model (blue dashed).
    The bottom panel shows the residuals between the models.
    The black vertical solid line marks the time of true pericenter passage,
    while the blue vertical dashed line marks the timed passage.
    These lines lie on top of each other as the difference between them is $0.018M$, or 0.001\% of the true $t_{\mathrm{p},1}-t_{\mathrm{p},0}$.}
  \label{fig:polarcomp}
\end{figure}

In Fig. \ref{fig:polarcomp} we compare our re-summed modes at the first subsequent
passage to the numerical integration using the initial conditions of our previous example,
now including radiation reaction in its evolution.
We give the re-summation the time and orbital parameters that
the numerical integration finds at the first passage, $t_{\mathrm{p},0}$,
and let the timing model (Eqs. \ref{eq:p_model} - \ref{eq:t_model})
calculate them at subsequent passages.
This model appears in blue, dashed lines.
The black line illustrates the numerical integration, and we can see that they align very well, the timing model offset by only $0.018M$.
This accuracy is despite the fact that the range of parameters used to fit the timing model did not include all the parameters of this orbit at $t_{\mathrm{p},0}$, $p\approx 9.77 M$ and $\eta \approx 0.100$, although it did include the eccentricity $e \approx 0.875$.
Thus, we have accurately modeled the excitation of f-modes across a pericenter passage.


\subsection{Waveform Match}
\label{sub:match}

\begin{figure*}[]
  \includegraphics[width=0.455\textwidth]{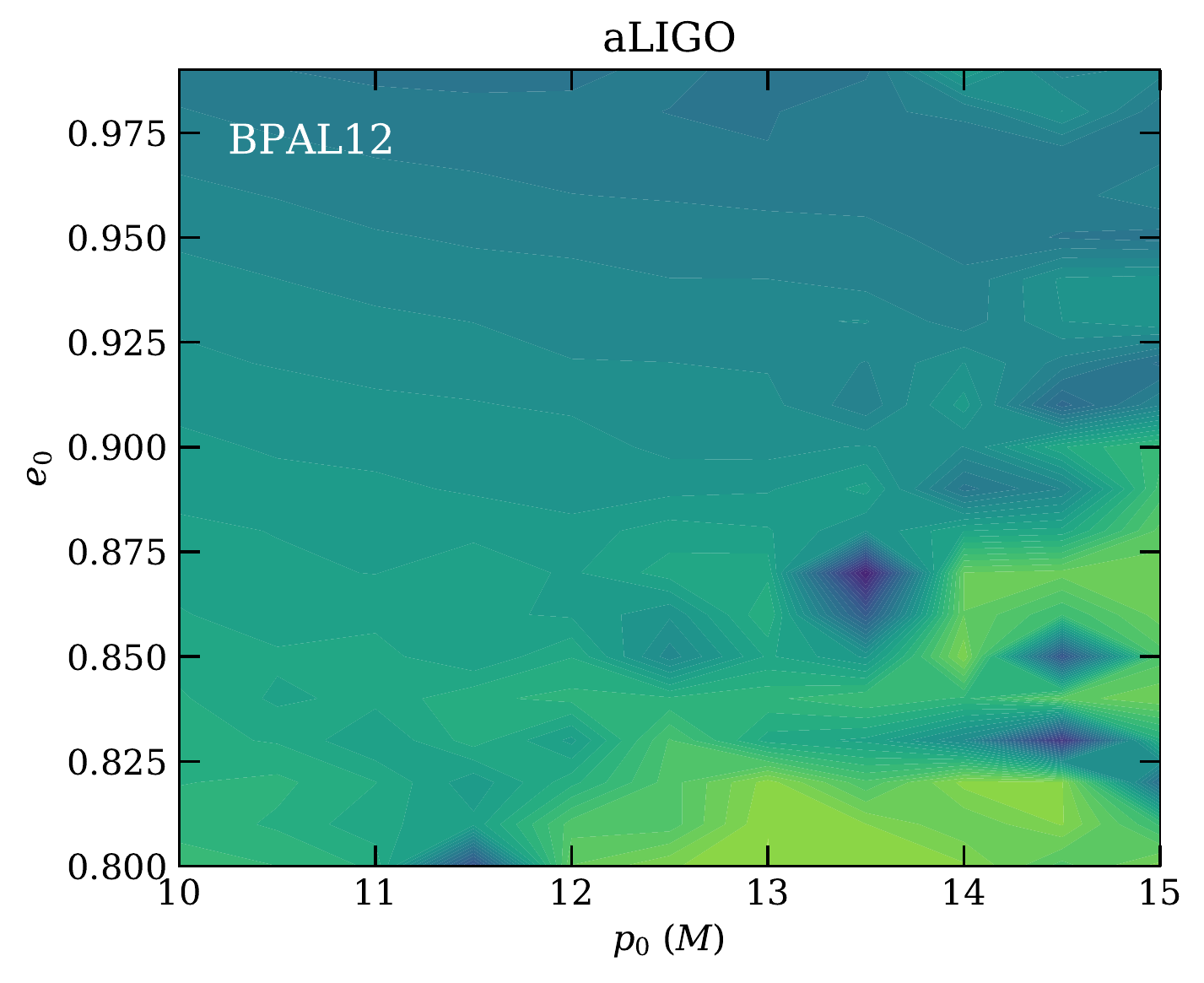}
  \includegraphics[width=0.535\textwidth]{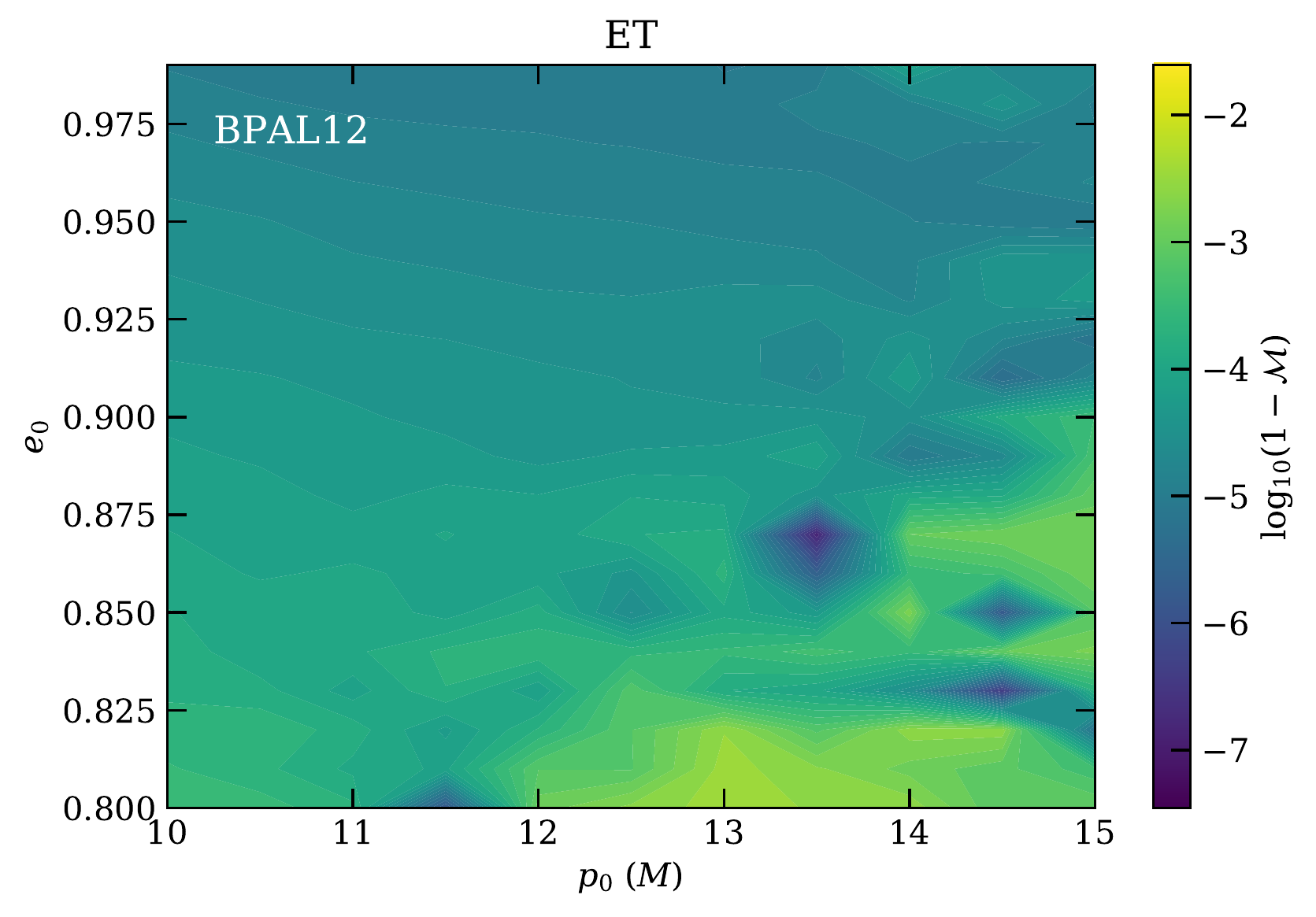}
  \includegraphics[width=0.455\textwidth]{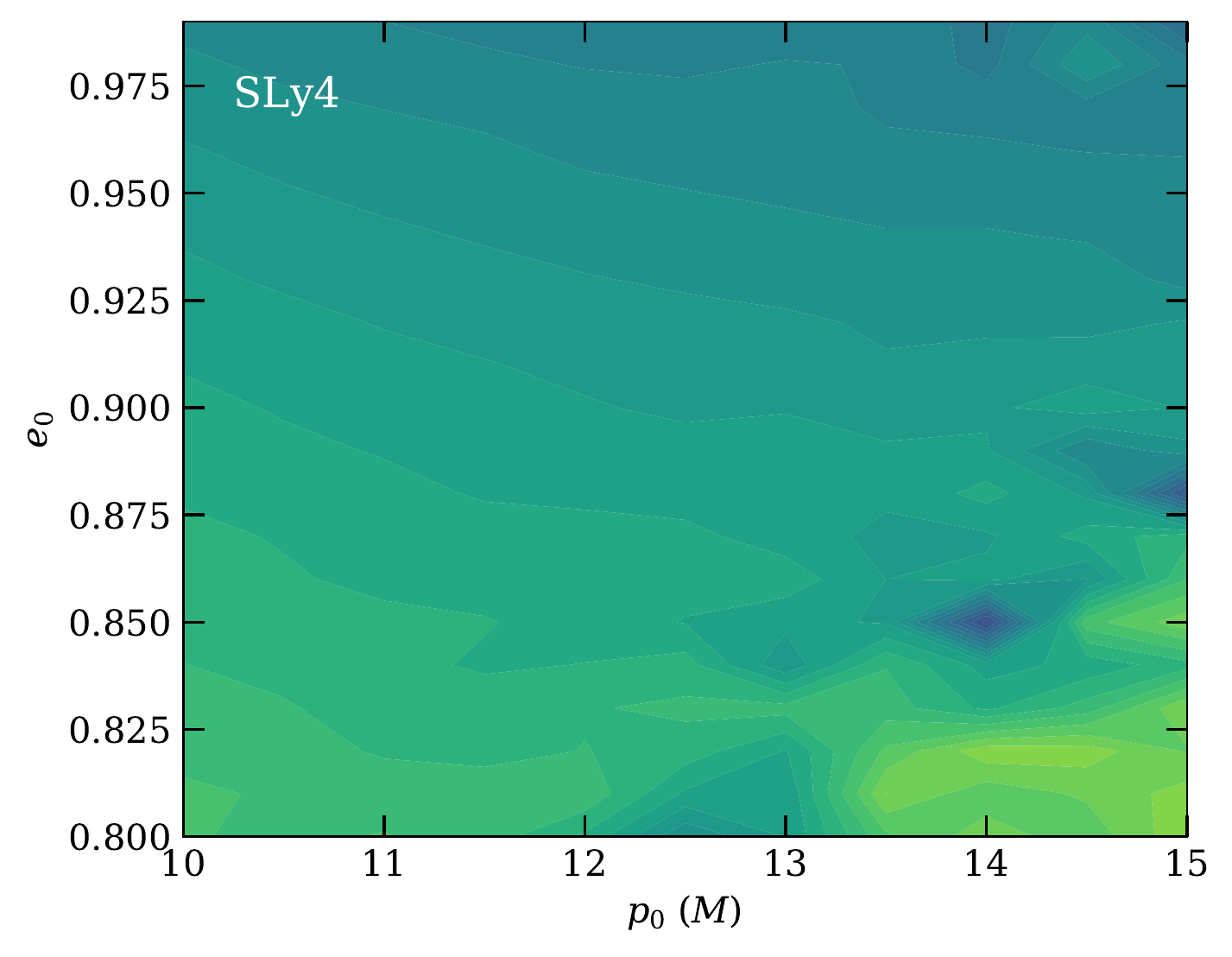}
  \includegraphics[width=0.535\textwidth]{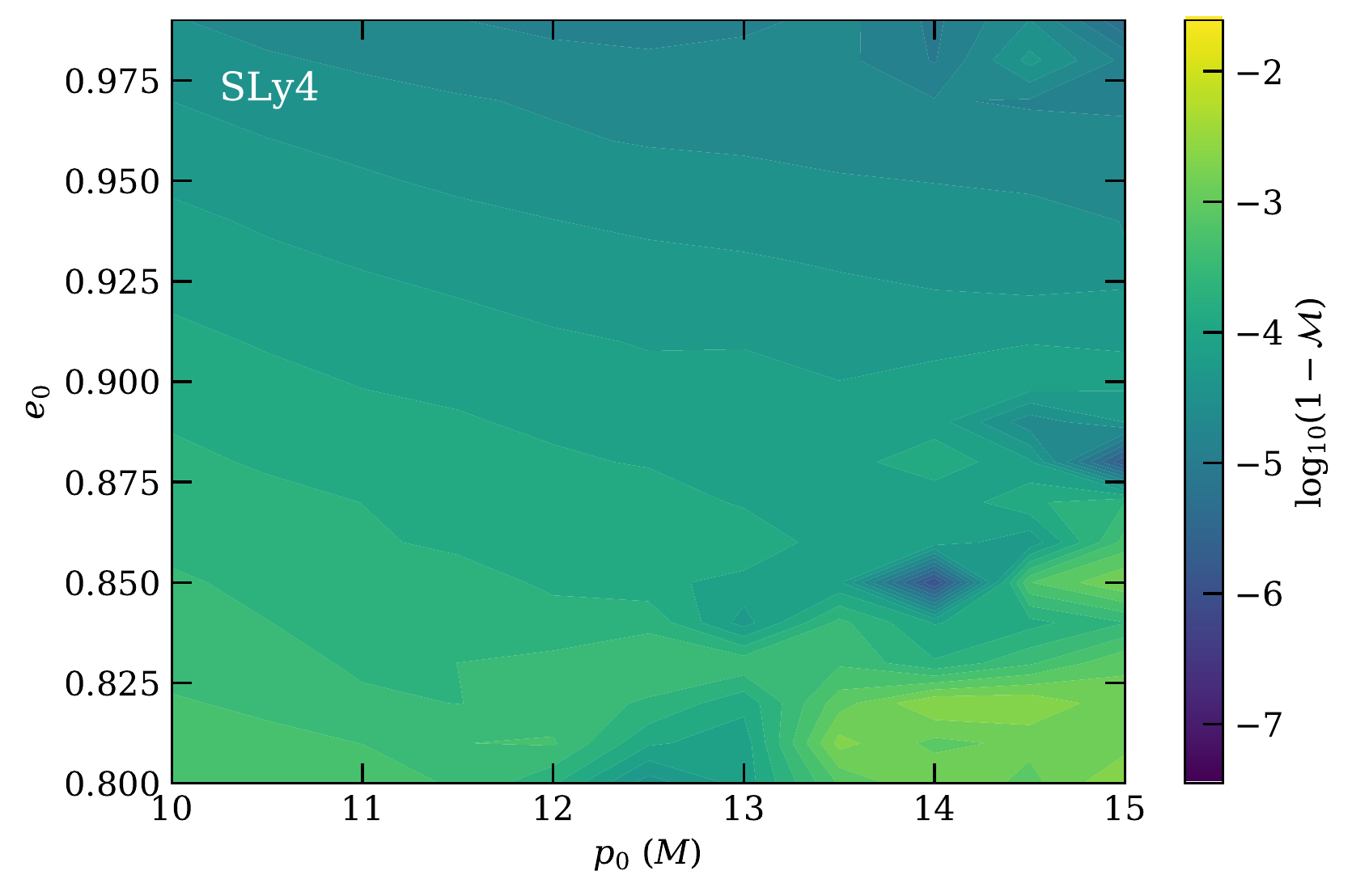}
  \includegraphics[width=0.455\textwidth]{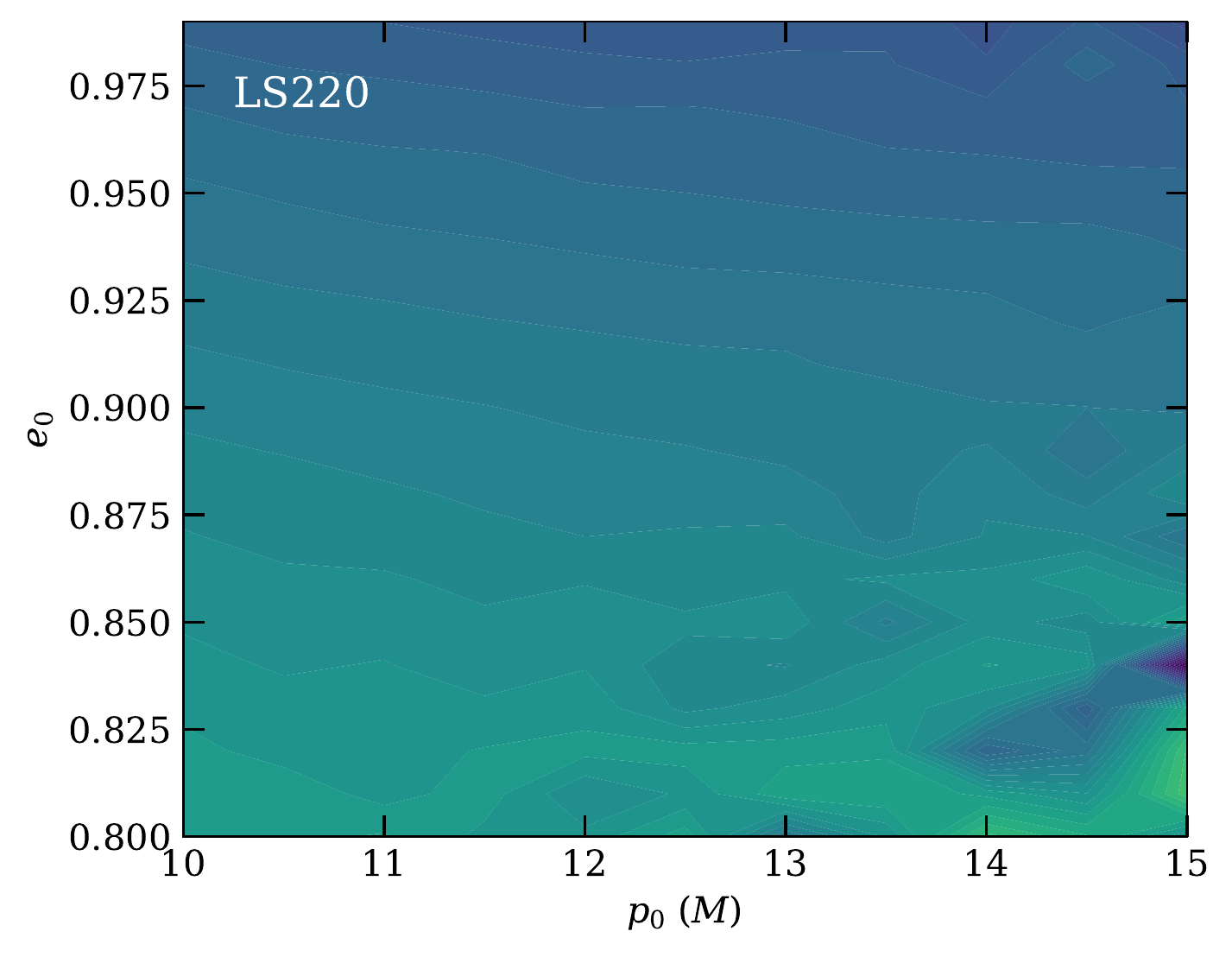}
  \includegraphics[width=0.535\textwidth]{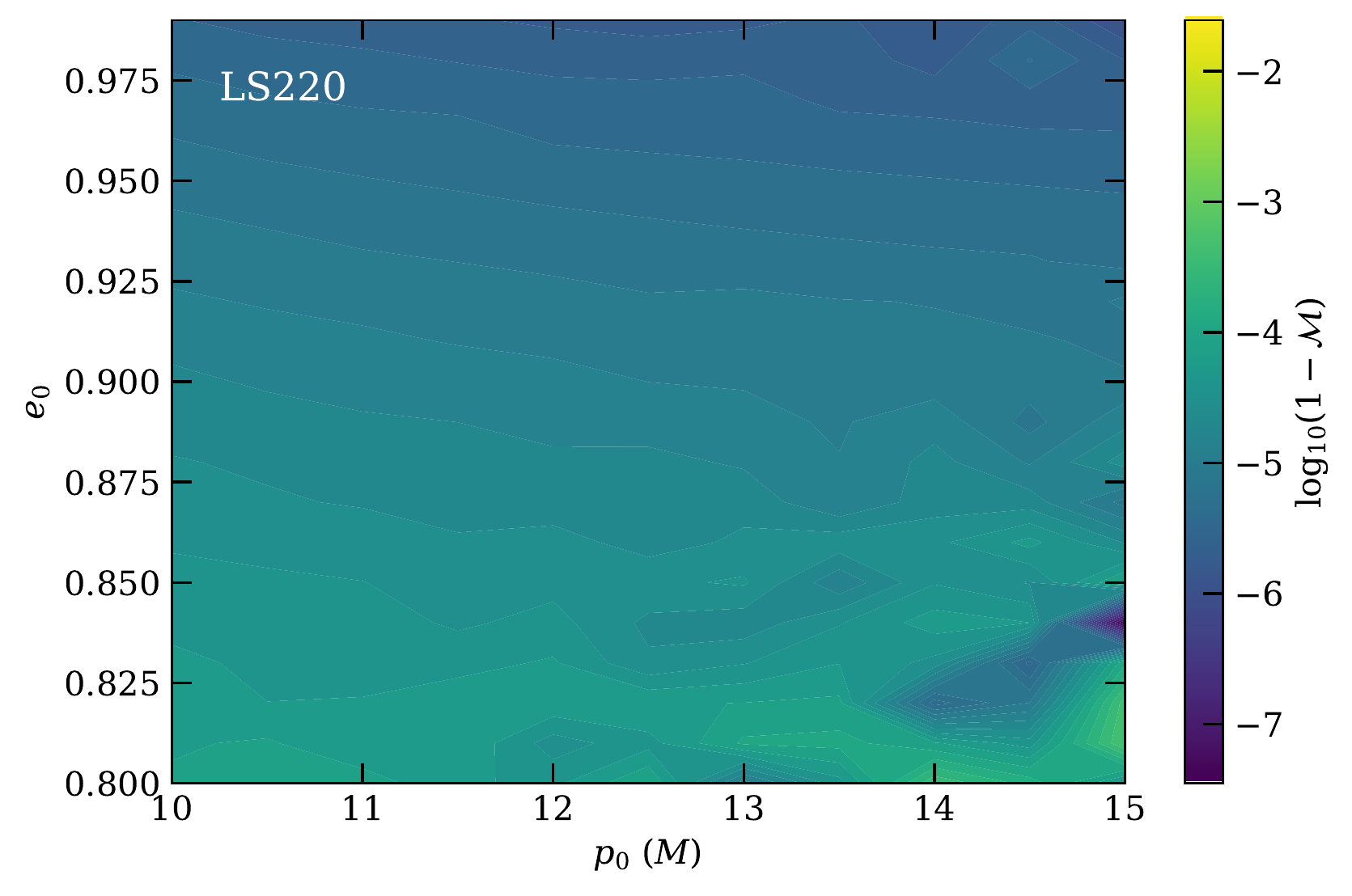}
  \caption{Mismatch between re-summed modes and numerical waveforms for BH--NS binaries as a function of $e$ and $p$ with $\MB = 10$ M$_\odot$ and, from top to bottom, NSs of $M_* = 1.223$ M$_\odot$ (BPAL12), $M_* = 1.273$ M$_\odot$ (SLy4 and LS220).
  	Their fundamental frequencies and decay timescales are, respectively, $f = 2.122 \textrm{ kHz}, \tau = 0.187 \textrm{ s}$; $f = 1.865 \textrm{ kHz}, \tau = 0.230 \textrm{ s}$; and $f = 1.628 \textrm{ kHz}, \tau = 0.306 \textrm{ s}$ \cite{Chirenti2015}.
    The mismatches in the left column were calculated with the aLIGO sensitivity curve
    while the mismatches in the right were calculated with ET.
    The color normalization is the same across all plots.}
  \label{fig:match}
\end{figure*}

In Sec. \ref{sub:comp}, we compared the time-domain waveforms evaluated
from numerical integration of the f-modes and from their re-summation
for a specific $e_0$ and $p_0$.
To estimate how well the re-summation would perform in detecting an f-mode
GW, however, one should vary these initial parameters to understand
where our fly-by approximation applies.
In this section, we do this across a range of parameters to calculate the match
of a single f-mode excitation between the re-summation and the numerical waveforms.

The match between the waveforms $h_{\rm A}$ and $h_{\rm B}$ is defined as
\begin{equation}
  \label{eq:match}
  \mathcal{M} = \underset{(\delta t,\delta\phi)}{\rm max}\frac{\left(h_{\rm A} \big| h_{\rm B}\exp[2\pi if \delta t + i\delta \phi]\right)}{\sqrt{\left(h_{\rm A} \big| h_{\rm A}\right) \left(h_{\rm B} \big| h_{\rm B}\right)}}
\end{equation}
where $\delta t$ and $\delta \phi$ are overall time and phase shifts,
and the noise-weighted inner product is
\begin{equation}
  \label{eq:inprod}
  \left(h_{\rm A} \big| h_{\rm B}\right) = 4 \text{Re} \int \frac{\tilde{h}_{\rm A}(f) \tilde{h}^*_{\rm B}(f)}{S_n(f)} \dee f
\end{equation}
where $\tilde{h}(f)$ is the Fourier transform of $h_+(t)$
and $S_n$ is the spectral noise density of the detector under consideration.
The match generally ranges from 0 ($h_{\rm A}$ and $h_{\rm B}$ are perfectly out of phase)
to 1 ($h_{\rm A}$ and $h_{\rm B}$ are in perfect agreement) \cite{Moore:2019xkm}.
Thus, the closer $\mathcal{M}$ is to 1,
the more accurately our re-summation models the numerical waveform.

Similarly as in Sec. \ref{sub:comp},
we generate the numerical (A) and re-summed (B) plus polarizations for orbits with a range
of parameters $e_0 \in [0.8,1.0)$ and $p_0/M \in [10,15]$
at a sample rate of $2^{14}$ Hz.
At these high eccentricities and small semi-latus rectums, the closest approach between the black hole and neutron star is small and excites the sharp tides that we aim to model.
To exclude the low frequencies from the first pericenter passage and excitations from subsequent passages, the numerical waveforms are trimmed to an interval $t_{\rm i} \leq t < t_{\rm f}$ where $t_{\rm i} = 300M$ and $t_{\rm f} = t_{\rm p,0} + 0.9(t_{\rm p,1}-t_{\rm p,0})$.
They are then padded on either side to have a total length of $2^{24}$,
and we finally perform their Fourier transformation with the SciPy module \verb|fft|.

To calculate the match, we use the publicly available Advanced LIGO \cite{Barsotti}
and ET-D high frequency configuration \cite{Hild2010id} sensitivity curves.
While both have a similar sensitivity in the decahertz range,
ET improves on LIGO's strain by an order of magnitude in the kilohertz.
This is the range in which f-modes oscillate,
and thus of importance for their detection.
While ET has yet to start construction,
our estimates here provide an outlook for the future of GW science.

Fig. \ref{fig:match} shows the mismatch $1-\mathcal{M}$ in log-scale for our
range of orbital parameters and for the equations of state (in order of increasing softness) BPAL12, SLy4, and LS220.
At pericenter passage, where the tide is excited,
$e$ and $p$ will remain close to their initial value.
The re-summation performs best at larger $e$ and smaller $p$,
reaching matches above 0.98 across our parameter space.
The more eccentric the orbit is, the sharper the tidal excitation is at pericenter;
the closer the bodies are at that point, the stronger the excitation is.
For lower $e$/larger $p$ orbits, the tidal excitation of the star also generates an adiabatic ``spike'' during pericenter passage, which contributes to the gravitational wave signal at frequencies lower than the f-mode frequency $\omega$. This can be seen from Fig.~\ref{fig:resonance}, which has power at values $k\sim 10 -100$ for that specific orbit. Since this effect is at lower frequency than the f-mode, it will dominate the inner product integrals in the match, which will cause it to deteriorate since the re-summed mode does not capture this effect.
To mitigate this effect, we have set the lower limit of the inner product \ref{eq:inprod} to 100 Hz.
Regardless, the largest amplitude f-modes are generated at high $e$/small $p$ where tidal effects are the strongest, and we have found through the match calculation that the re-summed mode is capable of accurately capturing these effects. It is worth noting that difference between the detectors is small in our parameter space, although the precision of the matches are limited by our sampling.
The match generally improves the softer the equation of state, the mismatch decreasing by an order of magnitude between BPAL12 and LS220, as the f-mode will be larger and easier to capture with our re-summation.

\section{Discussion}
\label{sec:discussion}

As GW observatories continue to detect binary mergers from astrophysical environments,
the likelihood increases of detecting a binary with signatures of
dynamical formation,
making it imperative to have waveform models of eccentric binaries at hand.
Here we have developed the first analytic waveforms for the f-modes from
highly eccentric BH--NS binaries using the effective fly-by framework.
At leading PN order, we solved the f-modes in harmonics of the
Keplerian orbit and re-summed them to obtain a damped harmonic oscillator
excited at closest approach.
Comparing the re-summed modes to their numerical integration,
we find that they are an accurate representation for highly eccentric
and close pericenter passages,
where the tides are excited sharply.
We have also outlined and shown the feasability of timing these excitations
and adding them coherently in sequence.
However,
this model is incomplete,
and further work remains to be done to provide a more comprehensive analysis.

In this study, we have primarily focused on BH--NS binaries. However, the results presented here are also applicable to BNSs in the following way. In such a case, f-modes would be generated on both NSs, being sourced by the tidal potential of the companion. To leading PN order, this tidal potential is given by the monopole terms of the gravitational potentials of the compact objects. Thus, the total f-mode response and their gravitational waves would simply be the sum of the f-modes from both components. At higher PN orders, the f-modes will contribute to the tidal potential, but the response will be suppressed by $v^{10}$. We thus expect the non-linear interaction between modes to be subdominant, and one can treat BNSs using the superposition of modes modeled as we have done so here.

An essential component of our model is the timing of pericenter
passages, which we showed in Sec. \ref{sub:comp} to be accurate
for close passages.
Nevertheless, work is needed on improving the timing model's accuracy beyond the effects of the 2.5PN radiation reaction.
The inclusion of tidal effects on the orbital evolution must be investigated,
as they hasten the decay of an orbit and introduce new physics into the timing
model.
We will pursue these tasks in a future study to further the model presented here.

A source of contention in the applicability of our model are the
Hansen coefficients, for which a simple closed form has eluded us.
As the indices of interest $k_\pm \sim \omega_m/n$ are of large magnitude,
one might expect the stationary phase approximation (SPA) to be useful in
evaluating Eq. \eqref{eq:fourier},
but the SPA does not account for $k_\pm$ having an imaginary component.
Resorting to evaluating the coefficients with numerical quadrature
has shown to take a large portion of the time spent calculating re-summed
modes, which makes this procedure less attractive for obtaining
models on the fly.
Tabulating $X^{-3,m}_k$ across values of $k$ may alleviate this concern,
but nonetheless holds back the model from being fully analytic.
Their form in infinite series has been utilized to 1PN order in \cite{Mikoczi:2015ewa},
but one must truncate them to a precision of their choice, balancing accuracy for evaluation time -- a consideration that has to be taken when applying these eccentric models.

In conclusion, we have only begun development of analytic waveforms
for highly eccentric binaries that include the effects of dynamical tides.
While the model can be improved by much,
we have shown how well our re-summation can match with numerical
waveforms for eccentric and close pericenter passages.
Such orbits can form in dense stellar environments,
and if their parameters are within our region of high match,
our re-summation would be a candidate for detecting them
and thus characterizing their formation channel.

\acknowledgments

N.L. acknowledges support from NSF grant PHY-1912171, the Simons Foundation, and the Canadian Institute for Advanced Research (CIFAR). We would like to thank Frans Pretorius for useful discussions.

\appendix
\section{Line-of-Sight Vector to NSs}
\label{app:los}

To project the metric perturbation due to the NS to the
transverse-traceless gauge,
we establish a tetrad defined by a line-of-sight vector
to an observer from the NS's center of mass,
$\bm{\hat{d}}_{*} =  \bm{D}_{*} / |\bm{D}_{*}|$,
where
\begin{equation}
  \bm{D}_{*} = d_{\rm L}\bm{\hat{d}} - \bm{R}_{*},
\end{equation}
and $\bm{\hat{d}}$ is defined in Eq. \eqref{eq:d} along with the
luminosity distance $d_{\rm L}$.
$\bm{R}_{*}$ is the vector from the binary center of mass
to the NS,
\begin{equation}
  \bm{R}_{*} = \frac{\MB}{M} R \left[\cos\phi, \sin\phi, 0\right],
\end{equation}
where $R$ and $\phi$ describe the binary orbit (see Sec \ref{sub:harmdecomp}).
We define two vectors perpendicular to this line-of-sight,
\begin{equation}
  \bm{y}_* = \bm{\hat{z}} \times \bm{\hat{d}}_{*}, \quad
  \bm{x}_* = \bm{y}_* \times \bm{\hat{d}}_{*}
\end{equation}
to complete the tetrad with the transverse unit vectors
\begin{equation}
  \bm{\Theta}_* = \bm{x}_*/|\bm{x}_*|, \quad
  \bm{\Phi}_* = \bm{y}_*/|\bm{y}_*|.
\end{equation}
Unlike $\bm{\hat{d}}$ for a stationary orbit (with respect to the observer),
these vectors are time-dependent,
describing the NS's revolution around the binary's center.
This modulates the f-mode's waveform on an orbital timescale.
However, the separations between the NS and the center of mass
we consider for tidal effects to be significant do not exceed
the hundreds of kilometers,
while the distances to binaries are expected to be
at least in the kiloparsecs if in our galaxy.
Thus, expanding this tetrad in $1/d_{\rm L}$,
we find that the leading order terms give Eqs. \eqref{eq:d}-\eqref{eq:Phi},
the same tetrad for the binary's center of mass.

\section{Fitting a Timing Model}
\label{app:timing}

We here discuss the procedure for developing the timing model described in Sec.~\ref{sub:timing}. Our starting point is the osculating equations in Eqs.~\eqref{eq:dpdt}-\eqref{eq:dvdt}. These equations can be converted to ordinary differential equations in $V$ by dividing by Eq.~\eqref{eq:dvdt} and PN expanding. For $(p,e)$, we obtain the equations
\begin{align}
\frac{\dee p}{\dee V} &= - \frac{16}{5} \eta \frac{M^{5/2}}{p^{3/2}} \left(1 + e \cos V\right) \left(4 + e^{2} + 5 e \cos V\right)\,,
\\
\frac{\dee e}{\dee V} &= - \frac{2}{15} \eta \left(\frac{M}{p}\right)^{5/2} \left(1 + e \cos V\right) \left\{\left(96 + 109e^{2}\right) \cos V 
\right.
\nn \\
&\left.
+ e \left[104 + 6 e^{2} + 2 \left(56+9e^{2}\right) \cos(2V) + 35 e \cos(3V)\right]\right\}\,.
\end{align}
To solve these, we propose the ansatz
\begin{equation}
\label{eq:ansatz}
	\mu^a = \sum_{j=0}^J \eta^j \mu^a_j
\end{equation}
for the orbital parameters $\mu^a = \{p,\epsilon\}$, and where we use $\eta$ as an order-keeping symbol due to the fact that the forcing function in the above equations scales linearly in $\eta$.
Inserting these equations into our $\dee\mu^a/\dee V$ and expanding them in $\eta$, we obtain evolution equations for each $j$-order.
As one expects, $\dee\mu^a_0/\dee V = 0$; this is the Newtonian condition that the orbital parameters be constants.

At first order in $\eta$, we obtain the solutions
\begin{widetext}
\begin{align}
\label{eq:p1}
p_{1}(V) &= -\frac{16}{5} \frac{M^{5/2}}{p_{0}^{3/2}} \left[\frac{1}{2} (8 + 7e_{0}^{2}) V + e_{0} (9+e_{0}^{2}) \sin V + \frac{5}{4} e_{0}^{2} \sin(2V)\right]
\\
\label{eq:e1}
e_{1}(V) &= - \frac{2}{15} \left(\frac{M}{p_{0}}\right)^{5/2} \left[\frac{1}{2} e_{0} (304 + 121 e_{0}^{2}) V + (96 + 269 e_{0}^{2} + 15 e_{0}^{4}) \sin V + 5 e_{0}(16 + 9 e_{0}^{2}) \sin(2V) 
\right.
\nn \\
&\left.
+ \frac{1}{3} e_{0}^{2} (91+9e_{0}^{2}) \sin(3V) + \frac{35}{8} e_{0}^{3} \sin(4V)\right]\,.
\end{align}
\end{widetext}
Evaluating these expressions at $V=2\pi$ gives us the recursion relations in Eqs.~\eqref{eq:p_model}-\eqref{eq:e_model}. 

To obtain the time of pericenter passage, one must integrate $\dee t/\dee V$, which is given by the reciprocal of Eq.~\eqref{eq:dvdt}. After PN expanding, the equation becomes
\begin{align}
\frac{\dee t}{\dee V} &= \frac{p^{3/2}}{M^{1/2}} \left(1 + e \cos V\right)^{-2} 
\nn \\
&- \frac{4}{15} \frac{\eta}{e} \frac{M^{2}}{p} \sin V \left(1 + e \cos V\right)^{-1} 
\nn \\
&\times \left[48 + 47 e^{2} + 2e (56+9e^{2}) \cos V + 35 e^{2} \cos(2V)\right]\,.
\end{align}
To integrate this equation, one must insert the ansatz in Eq.~\eqref{eq:ansatz}, expand in $\eta$, and apply the first order solution in Eqs.~\eqref{eq:p1}-\eqref{eq:e1}. Integrating from $V=0$ to $V=2\pi$ gives the time of pericenter passage, specifically
\begin{align}
\label{eq:t1}
t_{\rm peri} &= \frac{2\pi}{M^{1/2}} \left(\frac{p_{0}}{\epsilon_{0}}\right)^{3/2} 
\nn \\
&- \frac{2\pi^{2}}{5} \frac{M^{2}}{p_{0}} \left(\frac{425 - 366 \epsilon_{0} + 37 \epsilon_{0}^{2}}{\epsilon_{0}^{5/2}}\right) + {\cal{O}}(\eta^{2})\,,
\end{align}
where $\epsilon_{0} = 1 - e_{0}^{2}$. The first term in this expression in the orbital period of the unperturbed orbit, while the second is the correction to the orbital period from radiation reaction. Note that the correction is negative and diverges faster than the first as $e_{0}\rightarrow 1$, i.e. the second term scales as $\epsilon_{0}^{-5/2}$ while the first scales as $\epsilon_{0}^{-3/2}$. This implies that there is a region in parameter space where the second term can become larger than the first, and as a result, the time of the subsequent pericenter passage becomes negative, which is unphysical. This pathology is a result of the PN expansion that we have assumed in our ansatz.

To correct this pathology, we propose a re-summation of the time of pericenter passage of the form
\begin{equation}\label{eq:apptiming}
	t_{\rm per} = \frac{2\pi}{M^{1/2}} \left(
	\frac{p_0+\eta \left(\frac{M}{p_{0}}\right)^{5/2}A}{\epsilon_0 + \eta\left(\frac{M}{p_{0}}\right)^{5/2}B} \right)^{3/2}
\end{equation}
where we have chosen the functions
\begin{align}
	A &= a_0 + a_1 \epsilon_0, \\
	B &= b_0 + b_1 \epsilon_0 + b_2 \epsilon_0^2.
\end{align}
By PN expanding Eq.~\eqref{eq:apptiming}, one can obtain constraints on the parameters $(a_{0}, a_{1}, b_{0}, b_{1}, b_{2})$ by matching to the expression in Eq.~\eqref{eq:t1}, specifically
\begin{align}
\label{eq:b0-constraint}
b_{0} &= \frac{170\pi}{3}\,,
\\
a_{0} &= b_{1} + \frac{244\pi}{5}\,,
\\
\label{eq:a1-constraint}
a_{1} &= b_{2} - \frac{74\pi}{15}\,.
\end{align}
To obtain the full function, we must calibrate the free parameters $(b_{1},b_{2})$ against numerical integrations of Eq.~\eqref{eq:dpdt}-\eqref{eq:dvdt}.

To fit $(b_1,b_2)$, we generate data using the full PN equations \ref{eq:dpdt}-\ref{eq:dwdt} and calculate $t_{\mathrm{p},1}-t_{\mathrm{p},0}$.
This is done across the sets of parameters $\eta = \{1/4, 1/5, 1/6, 1/7, 1/8\}$, $p(t_{\rm p,0})/M = [10,20]$, and $e(t_{\rm p,0}) = [0.8, 0.994]$ to fit highly eccentric orbits.
We then use Mathematica's \verb|NonlinearModelFit| to find the best-fit values of the coefficients for each value of $\eta$ and $p(t_{\rm p,0})$;
that is, we fit explicitly for $t_{\mathrm{p},1}-t_{\mathrm{p},0}$ as a function of $e$. When performing the fitting, we found that using all of the constraints from Eqs.~\eqref{eq:b0-constraint}-\eqref{eq:a1-constraint} results in large residuals between the fitted model and the numerical data, typically a few hundred $M$. This is not sufficiently accurate for a timing model. Instead, we only use the constraint for $b_{0}$ and fit the remaining parameters using \verb|NonlinearModelFit|, which results in significantly smaller residuals.

\begin{figure*}[t]
\centering
\includegraphics[width=0.9\textwidth]{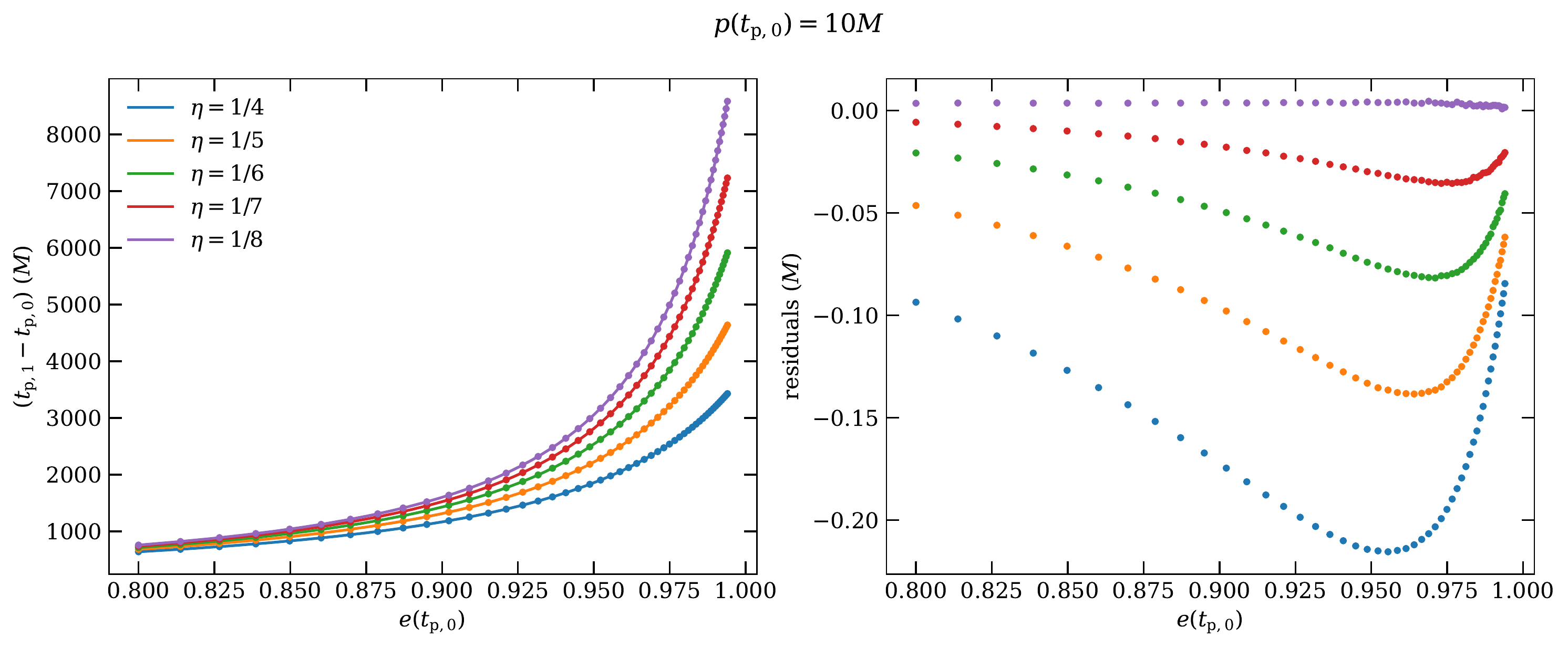}
\includegraphics[width=0.9\textwidth]{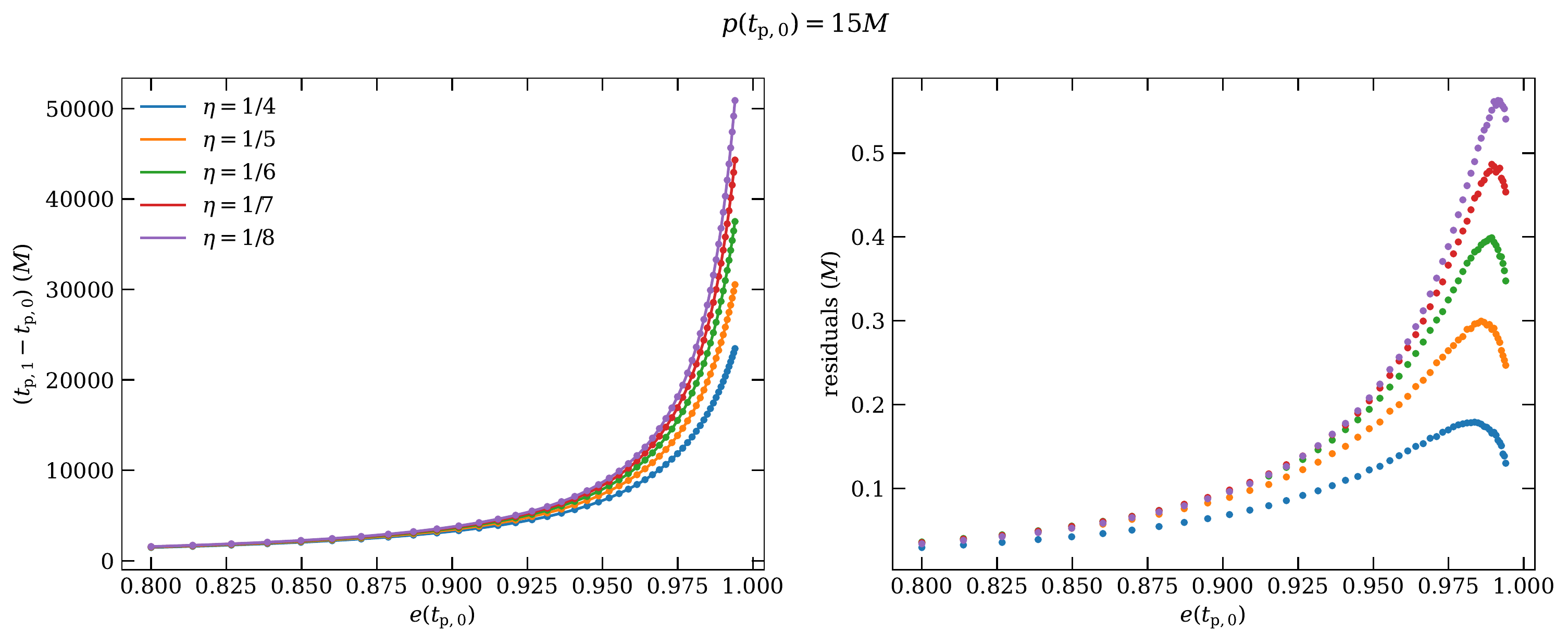}
\includegraphics[width=0.9\textwidth]{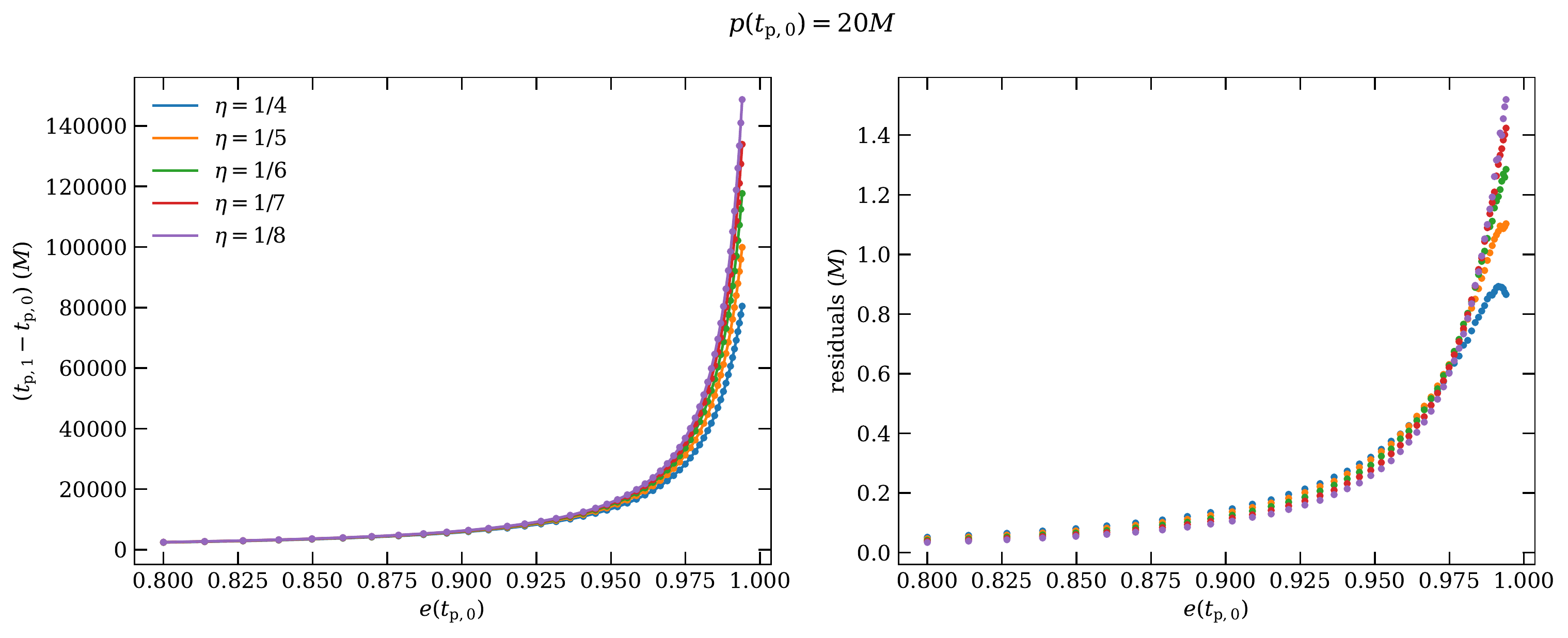}
\caption{\textit{Left column:} values of $t_{\mathrm{p},1}-t_{\mathrm{p},0}$ (colored dots) generated to fit Eq. \ref{eq:apptiming} and their resulting fits (solid lines); \textit{right column:} the respective residuals between our fitted model and the true time of pericenter passage.
	Each row is a set of values for fixed $p(t_{\rm p,0})$, where we show all the values for that semi-latus rectum for each $\eta$.
	In total we used 41 values of $p(t_{\rm p,0})$, but here we show only the first, median, and last of them to illustrate the model's accuracy concisely.
	While the residuals do grow with semi-latus rectum, they remain small for our parameter space.}
\label{fig:fits}
\end{figure*}

With values of $a_0$, $a_1$, $b_1$, and $b_2$ for different $\eta$ and $p(t_{\rm p,0})$, we looked for their trends as a function of these parameters.
The only coefficient with a clear trend was $a_0$, to which we then fit the model
\begin{equation}
	a_0 = a_{0,1} + a_{0,2} \left(\frac{\eta}{1/4}\right)^{a_{0,3}} \left(\frac{p}{10M}\right)^{a_{0,4}}.
\end{equation}
The remaining coefficients were taken to be their values averaged over $\eta$ and $p(t_{\rm p,0})$.
The results are listed in Eqs. \ref{eq:coeffs}-\ref{eq:a0coeffs}.
We illustrate the models and the generated data in Fig. \ref{fig:fits}, along with their residuals, which show that our model is accurate across our parameter space, its error at most on the order of $M$ for large semi-latus rectum, which suffices for our f-modes.

\bibliography{main}
\end{document}